\documentclass[%
reprint,
 amsmath,amssymb,
prb,
]{revtex4-2}
\usepackage{amssymb}
\usepackage{amsmath}
\usepackage{bbm}
\usepackage{bm}
\usepackage{epsfig}
\usepackage{color}
\usepackage{multirow}
\usepackage{graphicx}
\usepackage{booktabs}
\usepackage{dsfont}
\usepackage{tikz}
\usepackage{appendix}
\usepackage{makecell}
\usepackage{subfigure}
\usepackage{braket}
\usepackage{xcolor}
\usepackage[colorlinks,linkcolor=blue,anchorcolor=blue,citecolor=blue,urlcolor=blue]{hyperref}
\begin{document}

\title{Optimizing the Sensitivity–Noise Trade-off in Non-Hermitian Sensing via Off-Exceptional-Deficiency Operation}

\author{Shangxuan Li}
\affiliation{Department of Physics, Wuhan University of Technology, Wuhan 430070, China}

\author{Bin Guo}
\email{binguo@whut.edu.cn}
\affiliation{Department of Physics, Wuhan University of Technology, Wuhan 430070, China}

\date{\today}

\begin{abstract}
A central challenge in non-Hermitian sensing is that spectral singularities simultaneously amplify both the signal and environmental noise. We address this predicament in a double-chain Hatano-Nelson model featuring unidirectional interlayer coupling. At the exceptional deficiency (ED) limit, the system exhibits a macroscopically degenerate complex spectrum and a pronounced non-Hermitian skin effect (NHSE), yielding a sensitivity that scales exponentially with lattice size $N$ while remaining robust across a six-order-of-magnitude detuning range. By introducing diagonal spatial disorder, we demonstrate that the NHSE is progressively suppressed, whith eigenspace cosine similarity analysis quantifying a well-defined fault-tolerance threshold. To reconcile the sensitivity–noise trade-off, we delineate ”At-ED” and ”Off-ED” operating regimes. While the At-ED configuration imposes fractional-order noise amplification (SNR $\propto \delta^{-1/2}$) that saturates at a suboptimal plateau, migrating to the Off-ED regime eliminates this geometric singularity and restores a linear scaling law (SNR $\propto \delta^{-1}$), achieving an SNR enhancement of several orders of magnitude. Crucially, this improvement is achieved while fully preserving the exponential sensitivity scaling, albeit at a slightly reduced absolute sensitivity compared to the strict At-ED limit. Our findings establish the Off-ED framework as a concrete paradigm for next-generation topological sensors that reconcile extreme sensitivity with robust noise immunity.
\end{abstract}

\maketitle

\section{Introduction}
\label{sec:intro}

The field of non-Hermitian physics has substantially broadened our understanding of open quantum systems and dissipative phases of matter in recent years~\cite{Bender1998, Ashida2020, Okuma2023, Kawabata2019, ElGanainy2018, Gong2018}. Unlike conventional Hermitian systems, which strictly conserve probability and exhibit purely real eigenenergies, non-Hermitian Hamiltonians manifest a rich topological structure characterized by exceptional points (EPs)~\cite{Heiss2012, CanosValero2025, Bergholtz2021, Mandal2021, Tsuneya2026, Lin2025, Delplace2021, Xiao2021, Kononchuk2022, Kawabata2019a, Ding2016, MartinezAlvarez2018, Naghiloo2019, Ding2022} in the complex energy plane, where both eigenvalues and their corresponding eigenvectors coalesce simultaneously. In real space, non-Hermitian systems additionally give rise to the non-Hermitian skin effect (NHSE)~\cite{Okuma2020, Okugawa2020, Song2019, Yokomizo2021, Li2020a, Jiang2019, Zhang2021a, Zhang2021b, Kawabata2020a, Lin2023, Liang2022, Kawabata2023, Guo2021b, Okugawa2020, Phan2026, Claes2021, Gao2026, Rangi2025, Qin2025, Cai2026, Shu2025, Gliozzi2026, He2026, Yang2026, Li2026b, Yang2026a, Li2026c, Hu2025a}, a macroscopic boundary-localization phenomenon in which an extensive number of bulk eigenstates collapse onto the system boundaries under open boundary conditions (OBCs). This unconventional behavior marks a fundamental breakdown of the standard Bloch bulk-boundary correspondence~\cite{Kunst2018, Xiao2020, Yang2020, Nakamura2024, Zhang2026, Parasar2025, Hu2025b, Peng2025, Liu2026, Mardani2025, Mong2011, Rhim2018, Jin2019, Edvardsson2019, Hwang2019, Wang2020b, BayonaPena2025, Trifunovic2019}, thereby necessitating the framework of non-Bloch band theory~\cite{Yao2018, Song2019a, Yokomizo2019, Kaneshiro2025, Kawabata2020, Yokomizo2021a, Xue2021, Yokomizo2024, Verma2024, Hu2024, Wang2024b} to properly delineate the underlying topological phases.

By exploiting the higher-order degeneracies at EPs and the associated nonlinear spectral response to external perturbations (scaling as the $n$-th root of the perturbation strength $\epsilon^{1/n}$~\cite{Hodaei2017}), non-Hermitian topological sensors have been theoretically shown to possess sensing capabilities that surpass the fundamental limits of conventional Hermitian devices. In particular, sensor architectures that exploit the NHSE have been predicted to exhibit exponential sensitivity to boundary or parametric perturbations, owing to the directional amplification inherent in non-reciprocal hopping~\cite{Budich2020, Yuan2023}. Nevertheless, in realistic physical environments, non-Hermitian systems are inevitably subject to disorder arising from spatial impurities, fabrication imperfections, or fluctuating environmental backgrounds~\cite{Zhang2019a, Wiersig2020, Lau2018}. This introduces two compounding difficulties that severely hinder practical deployment. First, the extreme sensitivity stemming from the macroscopic accumulation of non-orthogonal eigenstates is typically accompanied by an exponentially narrow signal response window, which fundamentally restricts the dynamic range for linear detection~\cite{Koch2022}. Second, any global environmental noise or on-site fluctuations are strongly and indiscriminately amplified by the non-Hermitian gain mechanism, a detrimental behavior deeply linked to the divergence of the Petermann factor, leading to a severe deterioration of the overall signal-to-noise ratio (SNR)~\cite{Lee2008, Wang2020c}. Together, these issues constitute the central physical bottleneck obstructing the practical implementation of high-sensitivity non-Hermitian sensors.

Recently, exceptional deficiency (ED) has been proposed as a qualitatively new generalization of EPs to higher-dimensional state spaces~\cite{Li2026a, Bid2026}. In contrast to conventional EPs, which require stringent fine-tuning of system parameters and typically operate within an extremely narrow response bandwidth, ED systems exhibit broadly degenerate non-Hermitian dynamical evolution and anomalous coalescence behavior across a high-dimensional eigenspace. This opens up a promising avenue toward reconciling exceptional sensing sensitivity with multi-parameter robustness. Motivated by this potential, we incorporate this novel topological mechanism into a one-dimensional coupled double-chain Hatano-Nelson model. By engineering a highly asymmetric, unidirectional interlayer coupling between the two chains, the system can be driven precisely to the strict ED limit without disrupting the NHSE mechanism intrinsic to each individual chain. This setup uniquely exploits the intersection of real-space boundary localization and high-dimensional eigenspace coalescence.

While the concept of ED was introduced in Ref.~\cite{Li2026a} and its fundamental dynamical signatures were characterized therein, the present work addresses a distinct set of questions that are central to the practical deployment of ED-based sensors. Specifically, our contributions relative to Ref.~\cite{Li2026a} are threefold: (i) we provide the first systematic analysis of the sensitivity-noise trade-off within the ED framework, demonstrating that the geometric singularity responsible for exponential sensitivity simultaneously imposes a fractional-order noise amplification that caps the SNR; (ii) we identify the Off-ED regime as an optimal operating point that simultaneously preserves the exponential sensitivity scaling law and restores a linear SNR scaling (SNR~$\propto\delta^{-1}$), achieving an improvement of several orders of magnitude over the strict At-ED configuration; and (iii) we develop a channel-resolved decomposition of the global noise matrix, which pinpoints the left-boundary interlayer channel as the dominant noise gateway and identifies topologically immune, exponentially suppressed channels that can be exploited for targeted noise mitigation in realistic devices.

The remainder of this paper is organized as follows. In Sec.~\ref{sec:model}, we construct the real-space tight-binding Hamiltonian for the double-chain system and formulate the ED condition by applying the algebraic framework of generalized Jordan normal forms established in Ref.~\cite{Li2026a}. Sections~\ref{sec:eigenspace_coalescence} and~\ref{sec:disorder_robustness} examine the boundary-condition dependence of the complex energy spectrum, the spatial profile of the NHSE, and the coalescence dynamics of the multi-dimensional eigenspace induced by interlayer perturbations, as well as the system's robustness against on-site spatial disorder. In Sec.~\ref{sec:sensitivity_SNR}, we evaluate the key sensing metrics: we first demonstrate a broad amplification response window through explicit sensitivity calculations, and then, through a rigorous definition of the system noise matrix, establish the decisive advantage of operating in the off-ED regime for suppressing background noise fluctuations and maintaining an optimal SNR. Finally, Sec.~\ref{sec:conclusion} presents our conclusions and outlook.

\section{Double-Chain Hatano-Nelson Model and Exceptional deficiency}
\label{sec:model}

\begin{figure}[t!]
\centering
\includegraphics[width=0.4\textwidth,keepaspectratio]{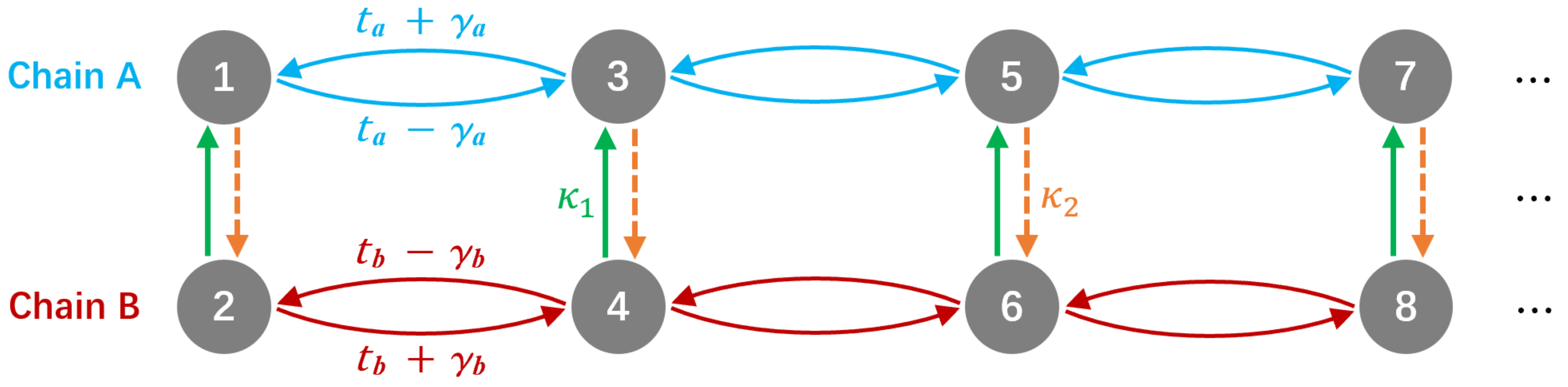}
\caption{Schematic of the double-chain Hatano--Nelson model with asymmetric interlayer coupling. Chain A (blue, upper) and chain B (red, lower) feature opposite nonreciprocal nearest-neighbor hopping with amplitudes $t_a \pm \gamma_a$ and $t_b \pm \gamma_b$, respectively. Solid and dashed arrows denote the dominant forward coupling $\kappa_1$ and reverse coupling $\kappa_2$, respectively, satisfying $\kappa_1 \gg \kappa_2$.}
\label{Fig1}
\end{figure}

To systematically investigate the degeneracy of a macroscopic high-dimensional eigenspace and the associated sensing response, we construct a non-Hermitian tight-binding system in real space consisting of two coupled parallel chains, the double-chain Hatano-Nelson model with unidirectional interlayer coupling, as schematically illustrated in Fig.~\ref{Fig1}. The system comprises two parallel non-Hermitian one-dimensional lattice chains (chain A and chain B), and its total Hamiltonian admits a natural decomposition into an unperturbed part and a perturbation term
\begin{align}
	H = H_0 + H_{\text{pert}}.
\end{align}

Within the tight-binding approximation and in the second-quantization representation, the full Hamiltonian is written as
\begin{align}
   H &= \sum_{n=1}^{N-1}\Bigl[(t_a-\gamma_a)a_{n+1}^\dagger a_n + (t_a+\gamma_a)a_n^\dagger a_{n+1} \nonumber\\
   &\quad + (t_b+\gamma_b)b_{n+1}^\dagger b_n + (t_b-\gamma_b)b_n^\dagger b_{n+1}\Bigr]\nonumber\\ 
   &\quad + \sum_{n=1}^{N}\bigl(\kappa_1 a_n^\dagger b_n + \kappa_2 b_n^\dagger a_n\bigr),
\end{align}
where $a_{n}^{\dagger} (a_n)$ and $b_{n}^{\dagger} (b_n)$ denote the fermionic creation (annihilation) operators acting on the $n$-th site of chain A and chain B, respectively. The model incorporates irreversible nonreciprocal intralayer hopping: $t_\xi$ ($\xi \in \{a, b\}$) represents the symmetric nearest-neighbor hopping amplitude, and $\gamma_\xi$ characterizes the strength of non-Hermitian asymmetry in the respective chain. Crucially, chains A and B are engineered to have opposite nonreciprocal hopping biases (as indicated by the inverted signs of $\gamma_a$ and $\gamma_b$ in the forward and backward hopping terms), so that each chain individually support intrinsic NHSE with opposing skin-localization orientations. The parameters $\kappa_1$ and $\kappa_2$ define the interlayer hopping rates from chain B to chain A and from chain A to chain B, respectively, enabling full control over the directionality and asymmetry of the interlayer coupling.

To solve for the eigenvalue dynamics and the ED of the system, we map the second-quantized operators onto a real-space basis, in which $H_0$ and $H_{\text{pert}}$ are rigorously expressed as $2N \times 2N$ block matrices
\begin{align}\label{Eq3}
	H_0 = \begin{pmatrix} H_A & \kappa_1 I \\  0 & H_B \end{pmatrix},\quad H_{\text{pert}}=\begin{pmatrix} 0 & 0 \\\kappa_2 I & 0\end{pmatrix}
\end{align}

A key feature of this construction is the deliberate imposition of a highly asymmetric interlayer coupling condition, $\kappa_1 \gg \kappa_2$. In the above block representation, $I$ is the $N \times N$ identity matrix, and $H_A$, $H_B$ are the $N \times N$ tridiagonal Hamiltonian matrices of the corresponding isolated single chains A and B, respectively, given explicitly by
\begin{align}
   H_A = \begin{pmatrix}
0      & t_a+\gamma_a & 0        & \cdots & 0        \\
t_a-\gamma_a & 0      & t_a+\gamma_a & \cdots & 0        \\
0      & t_a-\gamma_a & 0        & \ddots & \vdots   \\
\vdots & \vdots   & \ddots   & \ddots & t_a+\gamma_a \\
0      & 0        & \cdots   & t_a-\gamma_a & 0
\end{pmatrix},\\
H_B = \begin{pmatrix}
0      & t_b-\gamma_b & 0        & \cdots & 0        \\
t_b+\gamma_b & 0      & t_b-\gamma_b & \cdots & 0        \\
0      & t_b+\gamma_b & 0        & \ddots & \vdots   \\
\vdots & \vdots   & \ddots   & \ddots & t_b-\gamma_b \\
0      & 0        & \cdots   & t_b+\gamma_b & 0
\end{pmatrix}.
\end{align}

Combining the two contributions, the full Hamiltonian matrix $H$ takes the form of an off-diagonally asymmetric block structure
\begin{align}
	H = \begin{pmatrix} 
	H_A & \kappa_1 I \\  
	\kappa_2 I & H_B 
	\end{pmatrix}.
\end{align}

When the weak reverse interlayer coupling is completely eliminated, i.e., in the singular limit $\kappa_2 \to 0$, the system reaches the ideal ED regime described by the Hamiltonian $H_0$ in Eq.~\eqref{Eq3}, where the strictly unidirectional interlayer coupling drives the system into a defective eigenspace characterized by macroscopic Jordan-block deficiency. In this algebraic limit, $H_0$ exhibits a pronounced non-diagonalizable character. The absence of any operator pathway coupling eigenvectors from chain A back to chain B, manifested by the vanishing lower-left block, drives a severe geometric degeneracy of the eigenstates. Although the matrix has dimension of $2N$, its block-triangular structure leads to a proliferation of localized Jordan blocks. Specifically, for every degenerate eigenenergy $E$ shared by the two isolated chains, the unidirectional interlayer coupling $\kappa_1$ induces a Jordan block of the lowest possible dimension ($M=2$). While the maximum dimension of each individual Jordan block remains strictly bounded by $M=2$ across the entire spectrum, the total number of these defective blocks scales macroscopically with the lattice size $N$. This widespread clustering of low-order geometric degeneracies ensures that, in the singular limit $\kappa_2 \to 0$, the composite system loses exactly half of its linearly independent eigenvectors, thereby establishing the precise algebraic framework for the ED in this ladder structure.

The full set of conditions required for ED is as follows. First, under OBCs, the OBC spectra of the two isolated chains must be identical, i.e., $\eta_A = \eta_B$, ensuring complete spectral degeneracy between the two subsystems. Second, the block-triangular structure and spectral degeneracy alone are insufficient; the unidirectional interlayer coupling $\kappa_1$ must establish an effective connection between specific eigenstates of the two subspaces. Concretely, for every shared eigenenergy $E$, the matrix element $\langle v_A | \kappa_1 | u_B \rangle$ between the left eigenstate $\langle v_A |$ of $H_A$ and the right eigenstate $|u_B\rangle$ of $H_B$ must be nonvanishing~\cite{Li2026a}
\begin{align}
	\langle v_A|\kappa_1|u_B\rangle \neq 0.
\end{align}

This nonzero perturbative matrix element forces the corresponding right eigenstate $\ket{u_B} $to be mapped to the zero vector under the action of the full Hamiltonian, thereby collapsing the dimensionality of the associated eigenspace. The result is a defective Hilbert space populated by higher-dimensional Jordan blocks—the hallmark algebraic signature of ED.

According to the Hamiltonian parameters, the leftward hopping amplitude $(t_a+\gamma_a)$ within chain A exceeds the rightward amplitude $(t_a-\gamma_a)$, which necessarily drives the skin localization of chain A toward the left boundary (site index $1$). Conversely, chain B exhibits the opposite hopping asymmetry, where the rightward hopping amplitude $(t_b+\gamma_b)$ dominates over the leftward amplitude $(t_b-\gamma_b)$, thereby directing its intrinsic NHSE toward the right boundary (site index $N$). 

Under the ED construction with unidirectional interlayer coupling, the macroscopic skin direction of the composite system is determined by the chain whose nonreciprocal dynamics serve as the terminal sink of the unidirectional eigenstate flow. For the configuration in which the unidirectional interlayer coupling runs from chain B to chain A ($\kappa_1 \neq 0, \kappa_2 = 0$), the absence of any back-coupling pathway ($\kappa_2 = 0$) means that all eigenstate evolution terminates in chain A: the unidirectional flow irreversibly injects the eigenstates of chain B into the eigenspace of chain A, after which the nonreciprocal dynamics of chain A exclusively govern the spatial accumulation. Since chain A supports left-directed skin localization, all eigenstates accumulate macroscopically at the left boundary in real space. Conversely, when the unidirectional coupling runs from chain A to chain B ($\kappa_1 = 0, \kappa_2 \neq 0$), the terminal sink shifts to chain B: eigenstates are irreversibly injected from chain A into chain B, and the right-directed nonreciprocal dynamics of chain B then govern the global skin accumulation, giving rise to macroscopic accumulation at the right boundary. Both configurations exhibit a pronounced NHSE. Within the ED framework, the unidirectional interlayer coupling not only drives the spectral degeneracy of the composite system, but also acts as a terminal-sink selection mechanism that rigorously determines the global skin direction of the double-chain system by designating which chain serves as the irreversible endpoint of the eigenstate flow.

\section{Eigenspace Coalescence and Real-Space Distribution}
\label{sec:eigenspace_coalescence}

\begin{figure*}[t!]
\centering
\includegraphics[width=1\textwidth,keepaspectratio]{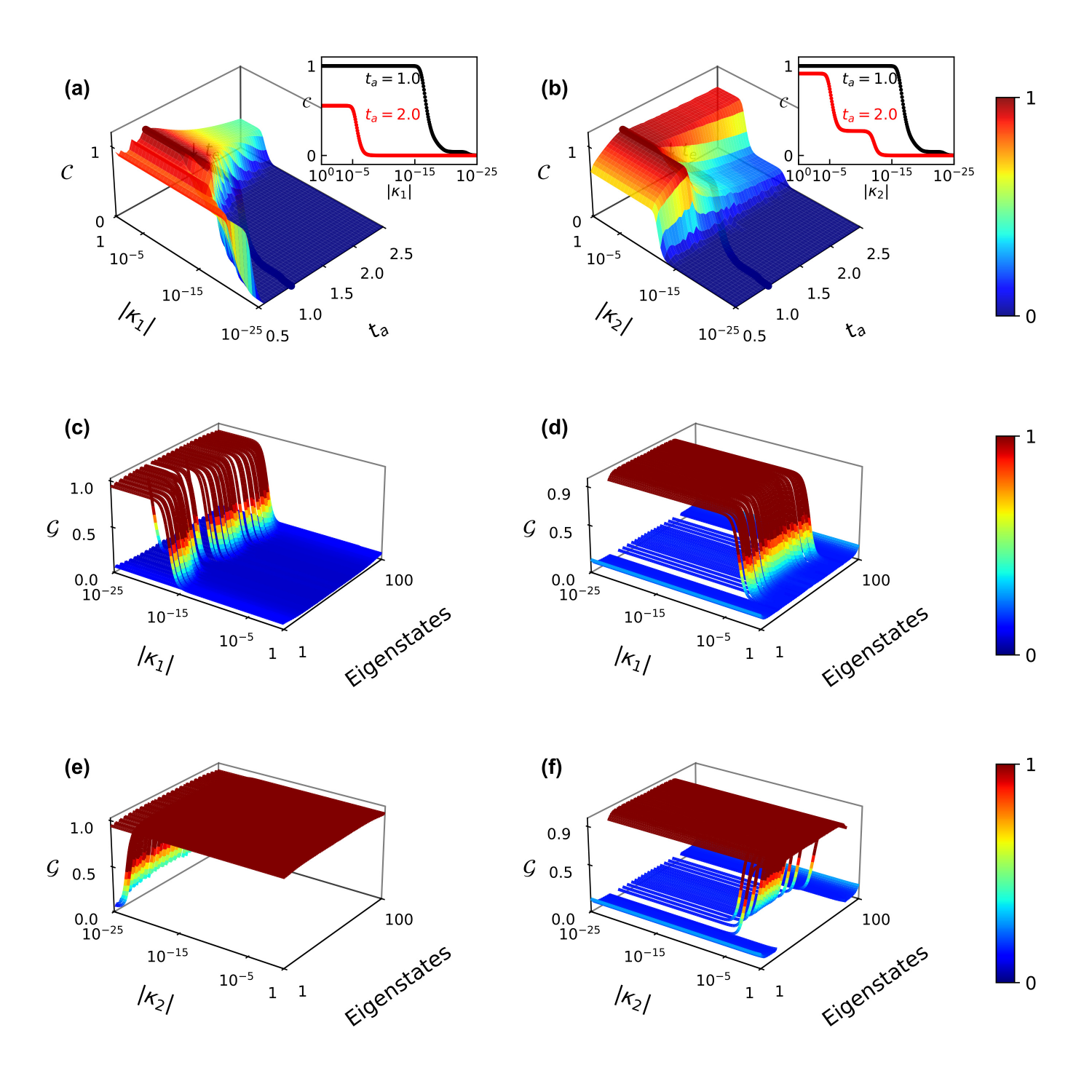}
\caption{Eigenspace coalescence and real-space skin-mode distribution under open boundary conditions. (a), (b) Cosine similarity $\mathcal{C}$ of the eigenspace as a function of interlayer coupling strength ($|\kappa_{1}|$ or $|\kappa_{2}|$) and hopping parameter $t_{a}$, for $B \rightarrow A$ and $A \rightarrow B$ coupling configurations, respectively. Insets display $\mathcal{C}$ versus $\kappa_{1,2}$ along two representative cuts at $t_a = 1.0, t_b= 1.0,\gamma_a = 0.5, \gamma_b = 0.5$ (black line, ED) and $t_{a}=2.0, t_b=1.0,\gamma_a=0.5,\gamma_{b}=0.25$ (red line, off-ED). (c), (d) Real-space spatial density $\mathcal{G}$ of all eigenstates as a function of $|\kappa_{1}|$ for B $\rightarrow$ A coupling. (e), (f) Corresponding spatial density $\mathcal{G}$ as a function of $|\kappa_{2}|$ for A $\rightarrow$ B coupling.}
\label{Fig2}
\end{figure*}

To rigorously quantify the coalescence of the eigenspaces of the double-chain system from the perspective of high-dimensional algebraic geometry, we introduce the subspace cosine similarity $\mathcal{C}(\epsilon(H_A), \epsilon(H_B))$~\cite{Li2026a}. This measure projects the multidimensional linear subspace $\epsilon(H_A)$ onto $\epsilon(H_B)$, thereby precisely evaluating the degree of non-orthogonal overlap between the two eigenspaces. Its algebraic form is defined via the Frobenius inner product and the Frobenius norm as
\begin{align}\label{eq:cosine_similarity}
    \mathcal{C}(\epsilon(H_A),\epsilon(H_B)) = \frac{\langle \epsilon(H_A), \epsilon(H_B) \rangle_F}{\|\epsilon(H_A)\|_F \, \|\epsilon(H_B)\|_F},    
\end{align}
where $\langle \cdot, \cdot \rangle_F$ denotes the Frobenius inner product and $\|\cdot\|_F$ is the Frobenius norm.

As shown in the insets of Figs.~\ref{Fig2}(a) and~\ref{Fig2}(b), under the strict ED condition ($t_a = 1.0, t_b= 1.0,\gamma_a = 0.5, \gamma_b = 0.5$, black line), the cosine similarity $\mathcal{C}$ maintains an absolute plateau at $\mathcal{C} = 1$ over an exceptionally broad range of coupling strengths $|\kappa_{1,2}|$. This indicates that the two independent eigensubspaces of the isolated chains have completely lost their orthogonality and linear independence, irreversibly coalescing into a single unified macroscopic eigenspace, a direct manifestation of higher-order Jordan-block formation. Conversely, under the off-ED condition ($t_{a}=2.0, t_b=1.0,\gamma_a=0.5,\gamma_{b}=0.25$), the cosine similarity $\mathcal{C}$ exhibits a sharp step-like decay upon moderate variations of $|\kappa_{1,2}|$ over several orders of magnitude. This behavior clearly demonstrates that the eigenspace coalescence away from the strict ED point becomes both incomplete and structurally fragile. Substantial values of $\mathcal{C}$ emerge only when the interlayer coupling becomes sufficiently large, a regime in which the two independent eigensubspaces begin to manifest significant overlap. For weak interlayer coupling, the negligible magnitude of $\mathcal{C}$ implies that these constituent eigensubspaces share almost no overlap, confirming that eigenspace assimilation and band coalescence become both incomplete and structurally fragile upon deviating from the ED limit. These results establish that only the strict ED condition can guarantee the absolute degeneracy and homogenization of the eigenspace at the macroscopic level, thereby forging a pure mathematical singularity. Crucially, this algebraic fragility implies that perfect eigenspace assimilation is highly susceptible to perturbations from even minute reverse interlayer coupling, providing an ideal geometric "switch" for regularizing the non-analytic response of the singularity while preserving the NHSE.

To quantitatively characterize the real-space distribution of eigenstates and the global modulation of the NHSE by interlayer coupling, we introduce the effective spatial density parameter $\mathcal{G}$ \cite{Li2026a}. As a physical observable measuring the spatial centroid of the OBC eigenstates, it is mathematically defined as
\begin{align} \label{eq:spatial_density}
    \mathcal{G} = \frac{1}{2N} \frac{\sum_{n=1}^{2N} n|\phi_n|}{\sum_{n=1}^{2N} |\phi_n|},
\end{align}
where $|\phi_n|$ is the amplitude of the eigenstate at site $n$. This definition provides a normalized measure of the spatial weight distribution across the lattice sites. By construction, $\mathcal{G}$ takes values in the interval $[1/(2N), 1]$: the minimum $\mathcal{G} =1/(2N)\approx 0$ is achieved when the entire weight is concentrated at site $n = 1$ (left boundary), while the maximum $\mathcal{G}=1$ is achieved when all weight resides at site $n = 2N$ (right boundary); a uniform distribution gives $\mathcal{G} \to (1+2N)/(4N)\approx 1/2$. Accordingly, when $\mathcal{G} \to 1/(2N)\approx 0$, the eigenstate is maximally localized at the left boundary, corresponding to left-directed non-Hermitian skin localization. When $\mathcal{G}\approx 1/2$, the eigenstate weight is approximately uniform across the bulk. When $\mathcal{G}\to 1$, the eigenstate is maximally localized at the right boundary, corresponding to right-directed skin localization.

Before turning to the numerical results, we briefly recall the terminal-sink selection mechanism established in Sec.~\ref{sec:model}. When the unidirectional coupling runs from chain B to chain A ($\kappa_1\neq 0, \kappa_2=0$), chain A serves as the irreversible endpoint of the eigenstate flow and its left-directed nonreciprocal dynamics govern the global skin accumulation, so that all eigenstates localize at the left boundary ($\mathcal{G}\to 0$). Conversely, for A $\rightarrow$ B coupling ($\kappa_1= 0, \kappa_2\neq 0$), chain B becomes the terminal sink and right-directed skin localization prevails ($\mathcal{G}\to 1$). Figures~\ref{Fig2}(c)–(f) numerically verify these predictions and further reveal how the NHSE competition evolves as the coupling strength departs from the ED limit. Crucially, the global skin direction is rigorously set by the dominant chain: we observe a clear left-boundary accumulation for B $\rightarrow$ A coupling and a right-boundary accumulation for A $\rightarrow$ B coupling. 

For the B $\rightarrow$ A coupling configuration, at the strict ED point ($t_a = 1.0, t_b= 1.0,\gamma_a = 0.5, \gamma_b = 0.5$), the spatial density $\mathcal{G}$ collapses onto a flat surface near its minimum value, unambiguously demonstrating that all OBC eigenstates of the composite system are governed entirely by the nonreciprocal dynamics of chain A, exhibiting absolute left-directed skin localization. As the system parameters deviate from the ED point (off-ED, e.g., $t_a = 2$, $\gamma_a = 0.5$, $t_b = 1$, $\gamma_b = 0.25$), the two chains progressively decouple with decreasing $\kappa_1$, and the real-space eigenstate distribution exhibits marked differentiation and a crossover region. Analogously, as shown in Figs.~\ref{Fig2}(e) and (f), for the A $\rightarrow$ B coupling configuration, $\mathcal{G}$ approaches unity throughout the spectrum in the ED limit, confirming that the global right-directed skin localization is entirely governed by the dynamical evolution of chain B. This systematic evolution of $\mathcal{G}$ demonstrates that the macroscopic spatial distribution of wavefunctions in non-Hermitian systems can be precisely engineered by tuning the unidirectional interlayer perturbation and the relative weight of competing skin modes.

\section{On-Site Spatial Disorder and Robustness Against Perturbations}
\label{sec:disorder_robustness}

\begin{figure}[t!]
\centering
\includegraphics[width=0.45\textwidth,keepaspectratio]{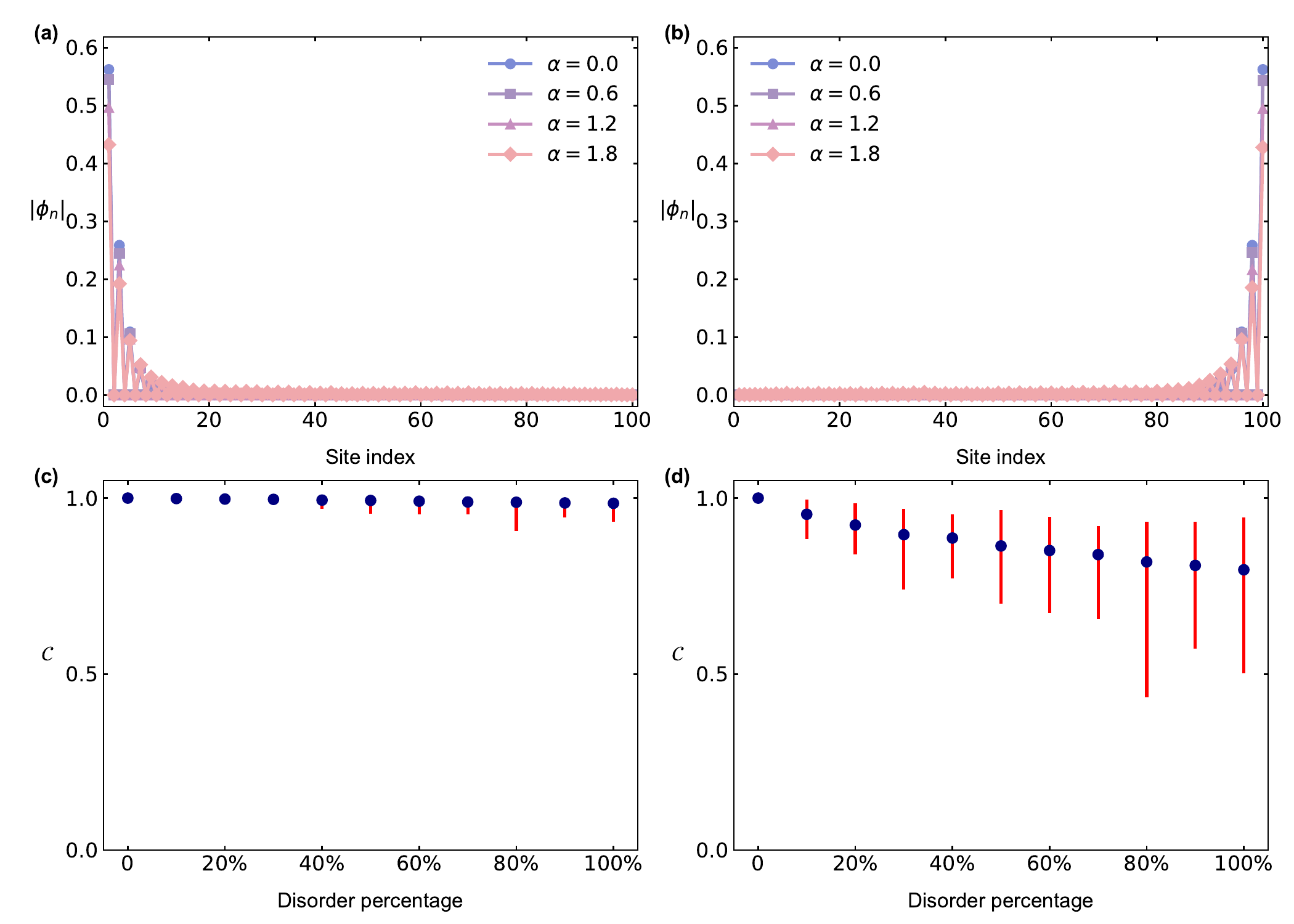}
\caption{Effects of on-site spatial disorder on eigenstate localization and eigenspace coalescence under open boundary conditions ($N=50$). (a), (b) Ensemble-averaged real-space probability density of the OBC eigenstates under varying disorder strengths $\alpha=0.0, 0.6, 1.2, 1.8$ for the $B\rightarrow A$ and $A\rightarrow B$ coupling configurations, respectively. (c), (d) Eigenspace cosine similarity $\mathcal{C}$ as a function of the disorder mask fraction $p$ for disorder strengths $\alpha=0.6$ and $\alpha=1.8$. Each data point (blue dot) represents the ensemble mean over multiple disorder realizations, with the red error bars indicating the corresponding standard deviation.}
\label{Fig3}
\end{figure}

Having established the pristine geometric properties of the ED state under pure boundary conditions, it is imperative to test its structural resilience against inevitable experimental spatial imperfections before evaluating its sensing capabilities.

In realistic experimental environments and device fabrication, the introduction of impurities and structural defects is unavoidable~\cite{Wiersig2020a, Landers2026, Loughlin2024, Zhong2019}. To rigorously examine the robustness of the non-Hermitian system against spatial disorder, we incorporate a spatially selective on-site random disorder perturbation into the real-space tight-binding Hamiltonian. Algebraically, this perturbation is strictly constructed as a purely diagonal $2N \times 2N$ matrix $D$, whose $i$-th diagonal element is defined as
\begin{align}\label{eq:disorder}
    d_i = m_i \cdot \omega_i,    
\end{align}
where $m_i \in \{0, 1\}$ is a real-space mask following a Bernoulli distribution with parameter $p$, which strictly dictates the spatial fraction of randomly perturbed sites within the physical lattice. The variable $\omega_i$ is sampled from a continuous uniform distribution over the interval $[-\alpha, \alpha]$, so that the on-site energy of each selected site undergoes a random shift within this range. In physical platforms such as photonic crystals, topological circuits, or acoustic metamaterials~\cite{Wang2019a, Feng2017, Yuan2023a, Xiao2019}, this diagonal perturbation corresponds to fabrication defects, material impurities, or local environmental fluctuations, which directly modify the resonant frequency of individual cavities or nodes, thereby completely breaking the spatial translational symmetry of the system and rendering the conventional Bloch theorem and momentum-space band theory inapplicable.

Nevertheless, the introduction of spatial disorder provides a direct test of the topological robustness of the ED state and the disorder resilience of the sensing performance. Under substantial fabrication imperfections (large $\alpha$ values), the NHSE continues to dominate the system dynamics, and the eigenstates remain preferentially localized at the lattice boundaries. This indicates that the nonreciprocal hopping exhibits considerable robustness against disorder, enabling the eigenstates to effectively overcome scattering by local impurities. The ensemble-averaged density profiles [Figs.~\ref{Fig3}(a),(b)] reveal that in the weak-disorder regime, the eigenstates remain robustly localized at the left (right) boundary, confirming the resilience of the NHSE against on-site perturbations. As $\alpha$ gradually increases, the boundary-localized distribution progressively broadens and extends gradually increases, the boundary-localized distribution progressively broadens and extends toward the lattice bulk. This evolution marks the onset of a weak-disorder immunity threshold, beyond which the randomized local potential scale begins to compete with the intra-chain asymmetric hopping. Distinct from standard localization phase transitions, this real-space diffusion phenomenon signifies that the double-chain system has reached the upper limit of its geometric fault-tolerance capacity. Under strong local perturbations, the coalescence of the high-dimensional eigenspace undergoes a systematic decoupling deviation, forcing the wavefunctions to transmute from a strictly boundary-clustered profile into a spatially extended configuration.

To further quantify the physical boundary of this disorder resilience from an algebraic-geometric perspective, we examine the evolution of the eigenspace cosine similarity $\mathcal{C}$ as a function of the disorder mask fraction $p$. As shown in Figs.~\ref{Fig3}(c) and~\ref{Fig3}(d), at a relatively weak disorder strength of $\alpha=0.6$, the cosine similarity $\mathcal{C}$ remains consistently close to unity irrespective of $p$. Crucially, for small $\alpha$ values, even under a full disorder mask fraction of $p=100\%$, $\mathcal{C}$ remains consistently close to unity with an exceptionally narrow standard deviation, explicitly confirming that the two eigensubspaces maintain a high degree of coalescence. In contrast, at a larger disorder intensity of $\alpha = 1.8$, $\mathcal{C}$ undergoes a pronounced degradation. These findings confirm that the eigenspaces of the two chains maintain a highly coalesced state under weak-to-moderate disorder perturbations, where the standard deviation remains essentially invariant, while under extremely severe disorder, both the degree of coalescence and its standard deviation exhibit conspicuous variations. These results not only underscore the robust fault-tolerance of the high-dimensional eigenspace coupling, but also explicitly substantiate, from the perspective of eigenspace projection, the exceptional resilience of the double-chain non-Hermitian system in complex disordered environments. In particular, the identification of a clear fault-tolerance threshold, below which the eigenspace cosine similarity $\mathcal{C}$ remains pinned near unity even at full disorder coverage ($p = 100\%$), establishes a well-defined operating window within which the ED-based sensing architecture retains its structural integrity.

\section{Sensitivity Scaling and Signal-to-Noise Ratio Optimization}
\label{sec:sensitivity_SNR}

\begin{figure*}[t!]
\centering
\includegraphics[width=1\textwidth,keepaspectratio]{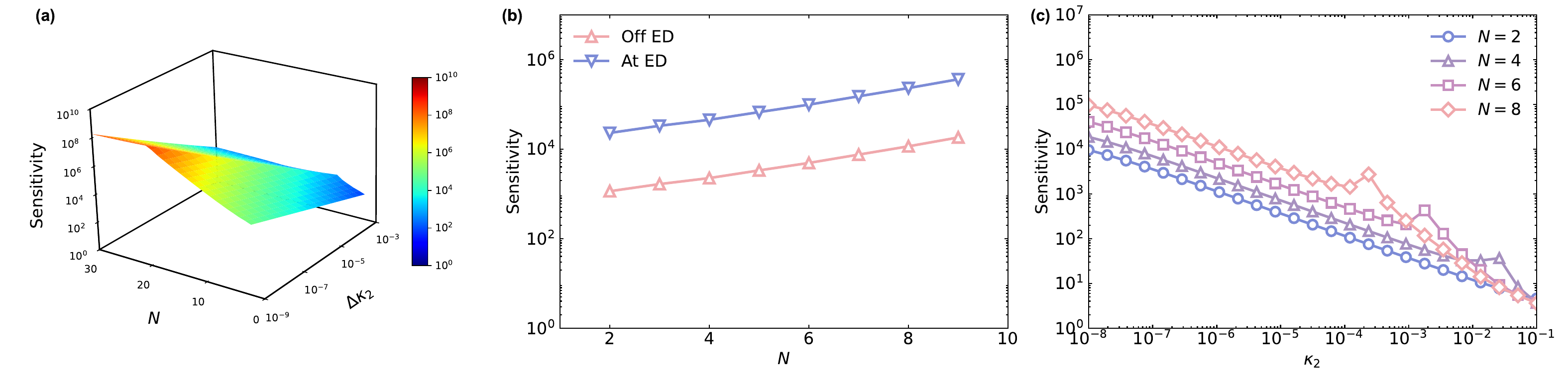}
\caption{Sensing performance of the double-chain Hatano-Nelson system. (a) Sensitivity $S$ as a function of perturbation strength $\Delta\kappa_{2}$ for various system sizes $N$. (b) Sensitivity $S$ versus lattice size $N$ at a fixed target signal $\Delta\kappa_{2}=10^{-8}$, comparing the At-ED and Off-ED ($\kappa_{2}=10^{-6}$) operating regimes. (c) Sensitivity $S$ as a function of background bias $\kappa_{2}$ for different system sizes $N$ in the Off-ED configuration. All results are obtained under open boundary conditions.}
\label{Fig4}
\end{figure*}

Having established in Sec.~\ref{sec:disorder_robustness} that the ED state maintains robust eigenspace coalescence under weak-to-moderate fabrication disorder ($\alpha < 0.6$ for $p = 100\%$), we henceforth under the assumption that on-site spatial disorder is kept within this fault-tolerance window and can therefore be treated as a controlled, static background rather than a dynamic noise source. It should be emphasized that the threshold $\alpha < 0.6$ characterizes the structural integrity of the eigenspace (as quantified by the cosine similarity $\mathcal{C}$), ather than the sensing performance directly; however, since the exponential sensitivity of the NHSE-based sensor is contingent on a well-coalesced eigenspace  ($\mathcal{C}\approx 1$), maintaining $\alpha < 0.6$ is a necessary prerequisite for the sensing analysis that follows. With spatial disorder thus controlled, we now evaluate the sensing metrics by focusing on the dominant and unavoidable noise sources in realistic non-Hermitian sensing platforms: stochastic on-site potential fluctuations, intra-chain hopping disorder, and interlayer cross-talk variations arising from global environmental fluctuations. To quantitatively evaluate the non-Hermitian sensing performance of the double-chain Hatano–Nelson system, we adopt a sensitivity metric rigorously defined via the eigenspectral splitting~\cite{Zhang2026a}
\begin{equation}\label{eq:sensitivity}
    S = \frac{\Delta E_{\kappa_2}}{\Delta \kappa_2},
\end{equation}
where the spectral splitting range induced by the perturbation $\kappa_2$ is defined as
\begin{align}\label{eq:spectral_splitting}
    \Delta E_{\kappa_2} = &\left[E_{\max}(\kappa_2) - E_{\max}(\kappa_2{=}0)\right]\nonumber\\
    &- \left[E_{\min}(\kappa_2) - E_{\min}(\kappa_2{=}0)\right].
\end{align}

As shown in Fig.~\ref{Fig4}(a), in the vicinity of the ED limit, the sensitivity $S$ exhibits pronounced exponential growth with lattice size $N$ prior to saturation, explicitly demonstrating that the divergent scaling is fundamentally constrained by the intrinsic non-Hermitian dynamics of the system. Notably, the absolute value of $S$ exhibits a strong inverse dependence on the perturbation magnitude $\Delta\kappa_2$: a larger perturbation yields a smaller spectral response sensitivity. This observation reveals a fundamental physical trade-off in non-Hermitian supersensing, while the system can achieve exponentially amplified sensitivity, the corresponding signal response window is necessarily exponentially narrow. Nevertheless, as illustrated in Fig.~\ref{Fig4}(a), the detectable perturbation range spans a broad physical window covering six orders of magnitude, from $10^{-9}$ to $10^{-3}$. This expansive operational span underscores a remarkable detuning tolerance margin across multiple decades, over which the system robustly sustains continuous macroscopic signal amplification without requiring meticulous parameter fine-tuning. 

To maintain scientific rigor, we clarify that this expansive operational span should be construed as a remarkable "detuning tolerance margin" rather than a conventional linear response window. As the perturbation $\Delta\kappa_2$ escalates toward $10^{-3}$, the sensitivity curve exhibits a distinct downward gradient, signifying a transition from the strict linear perturbation regime into a nonlinear saturation zone. Crucially, unlike conventional EP sensors~\cite{Chen2017a, Hokmabadi2019, Mao2023}, where even minuscule parameter deviations trigger a catastrophic collapse in amplification performance, the ED architecture proposed here circumvents this limitation. Across the entire six-decade range of parameter mismatch, although the sensitivity degrades due to nonlinear saturation, its value remains robustly sustained within an immense positive interval (from $10^2$ to well over $10^8$). This substantiates the capability of the ED framework to yield exceptional, continuous macroscopic signal amplification that is independent of meticulous fine-tuning, thereby effectively decoupling extreme topological sensitivity from the rigid constraints of operational parameter fragility.

Based on the sensitivity metric defined above, we systematically compare the sensing performance in the strictly at-ED and off-ED regimes. As shown in Fig.~\ref{Fig4}(b), where the sensitivity $S$ is evaluated against the system size $N$ under a fixed target signal of $\Delta\kappa_{2}=10^{-8}$, the At-ED sensitivity consistently exceeds that of the Off-ED configuration (set at a background bias of $\kappa_{2}=10^{-6}$). Crucially, as the background bias $\kappa_{2}$ systematically decreases toward zero, the Off-ED sensitivity curve asymptotically converges to the strict ED limit, confirming that the At-ED configuration serves as the ultimate performance upper bound for this double-chain architecture. Importantly, although the absolute sensitivity value in the Off-ED regime is lower than the At-ED upper bound, the exponential scaling law $S \propto e^{\alpha N}$ remains fully intact throughout the Off-ED window ($\kappa_{2}<\kappa_2^c$), as evidenced by the parallel slopes of the two curves in the semi-logarithmic plot of Fig.~\ref{Fig4}(b). It is therefore precise to state that the Off-ED operation retains the exponential scaling while trading a marginal reduction in prefactor for a decisive improvement in SNR.

As illustrated in Fig.~\ref{Fig4}(c), the sensitivity $S$ is mapped as a function of the background bias $\kappa_{2}$ across varying lattice scales. This detailed landscape reveals a distinct topological restructuring threshold: once $\kappa_{2}$ exceeds a critical value, the reverse coupling immediately percolates across the entire physical lattice, triggering a dramatic global delocalization of the originally boundary-localized skin states. At this topological restructuring transition, the energy spectrum becomes extremely fragile and exhibits an anomalously divergent response to the background bias $\kappa_2$, manifesting as sharp spikes in the sensitivity curves. Beyond this critical threshold, the sensitivity decreases monotonically with increasing $\kappa_{2}$; remarkably, the sensitivity curves for all $N$ collapse onto a single universal profile regardless of system size. This scaling collapse signifies that the macroscopic NHSE is completely suppressed and the size-dependent non-Hermitian amplification advantage is entirely lost. Furthermore, reflecting a hallmark of finite-size scaling, the critical threshold shifts to larger values of $\kappa_2$ as $N$ decreases. Therefore, the background bias $\kappa_2$ must be judiciously maintained below this topological restructuring threshold to prevent the delocalization of skin modes and the associated degradation of sensing performance.

In any realistic physical measurement, the ultra-high sensitivity enabled by non-Hermitian singularities is fundamentally bounded by the simultaneous amplification of background fluctuations. To rigorously quantify the true noise-resilient sensing capability of our double-chain architecture, we evaluate the SNR defined as~\cite{Zhang2026a}
\begin{equation}
    \text{SNR} = \frac{\Delta E_{\text{sig}}}{\Delta E_n},
    \label{eq:SNR}
\end{equation}
where $\Delta E_{\text{sig}}$ represents the maximum eigenspectral splitting induced solely by the target perturbation $\Delta\kappa_2$ in the reverse interlayer channel. The denominator $\Delta E_n$ characterizes the collective spectral footprint of environmental noise, quantified via the deformation of the complex energy boundaries
\begin{equation}
    \Delta E_n = \left[E_{\max}(\delta) - E_{\max}(\delta{=}0)\right] - \left[E_{\min}(\delta) - E_{\min}(\delta{=}0)\right].
    \label{eq:noise_splitting}
\end{equation}

This noise-induced spectral response is generated by an all-encompassing global noise matrix spanning all spatial degrees of freedom
\begin{equation}
    \boldsymbol{\Omega}_{2N \times 2N} = \delta\left(\mu \mathbf{J} + \sigma \boldsymbol{\Lambda}\right),
    \label{eq:noise_matrix}
\end{equation}
where $\delta$ scales the overall noise amplitude, $\mathbf{J}$ is the all-ones matrix representing a uniform background fluctuation bias, and $\boldsymbol{\Lambda}$ is a $2N \times 2N$ random matrix with independent standard normally distributed entries~\cite{Mehta2004}. Consequently, each individual element of $\boldsymbol{\Omega}_{2N \times 2N}$ follows a Gaussian distribution with mean $\mu$ and standard deviation $\sigma$. Physically, this comprehensive noise architecture models the concurrent impact of on-site potential fluctuations, intra-chain hopping disorder, and cross-talk variations across the interlayer pathways ($\kappa_1, \kappa_2$)~\cite{Loughlin2024, Hatano1996}. By embedding stochastic perturbations across all matrix channels, the global formulation in Eq.~\eqref{eq:noise_matrix} provides a physically rigorous, unconstrained criterion for assessing the ultimate sensing capacity of macroscopic non-Hermitian systems under severe eigenspace aggregation.

\begin{figure}[t!]
\centering
\includegraphics[width=0.45\textwidth,keepaspectratio]{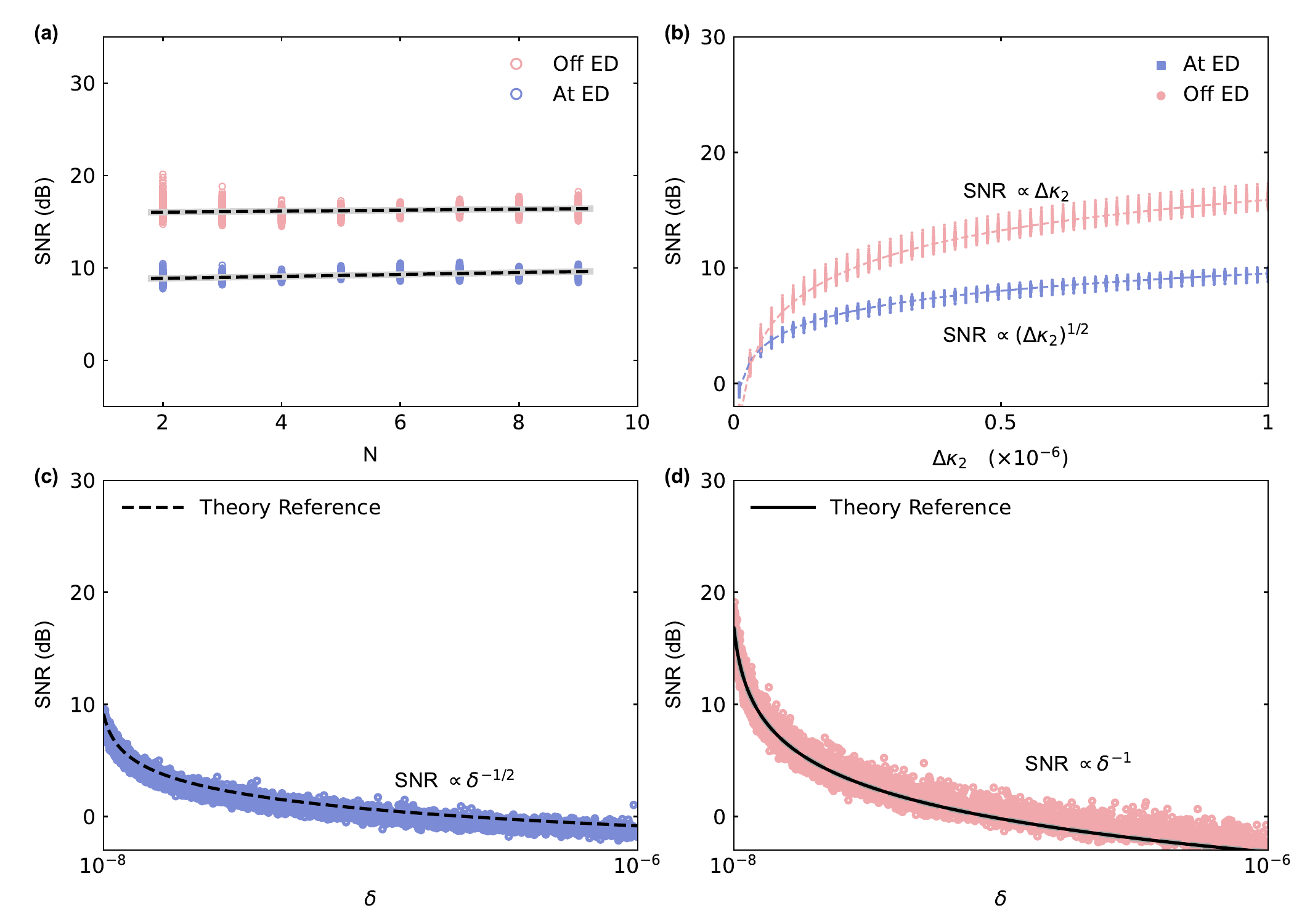}
\caption{Signal-to-noise ratio (SNR) comparison between the At-ED and Off-ED operating regimes. (a) SNR as a function of system size $N$ at a fixed noise level $\delta=10^{-8}$ for the At-ED and Off-ED ($\kappa_{2}=10^{-6}$) configurations, where dashed lines indicate the respective ensemble means. (b) SNR as a function of the target signal strength $\Delta\kappa_{2}$ at $\delta=10^{-8}$. (c) SNR as a function of noise level $\delta$ at the strict ED point, with the dashed line showing the theoretical $\delta^{-1/2}$ scaling. (d) Corresponding SNR versus $\delta$ in the Off-ED regime, with the solid line showing the theoretical $\delta^{-1}$ scaling.}
\label{Fig5}
\end{figure}

Following this theoretical framework, we conduct a systematic comparison of noise-resilient sensing performance between the At-ED and Off-ED regimes. As shown in Fig.~\ref{Fig5}(a), the SNR remains nearly independent of the lattice size $N$, indicating robust scale-invariant stability. Moreover, at a fixed perturbation strength of $\delta = 10^{-8}$ against a stable target signal source with $\Delta\kappa_2 = 10^{-6}$, the Off-ED regime consistently achieves a higher SNR than the At-ED regime, demonstrating its superior sensing performance under realistic noise conditions.

The apparent contrast between Figs.~\ref{Fig5}(a) and \ref{Fig5}(b) originates from their different parameter cross-sections. Figure~\ref{Fig5}(a) corresponds to a fixed perturbation $\delta = 10^{-8}$, which, as indicated by the dashed vertical line in Fig.~\ref{Fig5}(b), lies beyond the At-ED/Off-ED crossover point. Consequently, the Off-ED configuration operates within its linear-response regime, where the signal grows proportionally to $\Delta\kappa_2$ and yields a larger SNR. In contrast, Fig.~\ref{Fig5}(b) reveals the complete crossover behavior. For sufficiently weak perturbations below the crossover threshold, the At-ED regime temporarily exhibits a higher SNR by virtue of its intrinsic square-root response. As the perturbation strength increases beyond the crossover point, however, the Off-ED regime surpasses the At-ED regime and maintains a higher SNR over a broad parameter range. These observations consistently demonstrate that the sensing advantage of the At-ED regime is restricted to an ultra-weak perturbation window, whereas the Off-ED regime provides superior noise-resilient performance under experimentally relevant conditions.

Furthermore, Fig.~\ref{Fig5}(b) maps the SNR against varying target signal strengths $\Delta\kappa_{2}$. From an experimental standpoint, choosing the reverse interlayer variation $\Delta\kappa_{2}$ as the target sensing signal carries profound physical relevance for realistic device operations. In prominent non-Hermitian platforms such as photonic crystal waveguide lattices or topological circuit networks, the ideal unidirectional interlayer coupling ($\kappa_2 = 0$) is typically established using active nonreciprocal elements such as optical isolators or operational amplifiers. When the sensing architecture is exposed to external physical targets, such as micro-magnetic fields, localized mechanical stress, or ambient refractive index fluctuations, these external perturbations inevitably degrade the nonreciprocal isolation and induce a directional cross-talk leakage. The parameterized variation $\Delta\kappa_{2}$ precisely captures and quantifies the strength of this perturbation-induced directional leakage, thereby acting as a high-fidelity proxy for the external physical signals. In the weak-signal regime, the At-ED configuration exhibits a characteristic square-root scaling profile, $\text{SNR} \propto (\Delta\kappa_{2})^{1/2}$, which reflects the underlying square-root singularity of the coalesced eigenspace. As $\Delta\kappa_{2}$ scales up, the SNR at the strict ED point surges abruptly before plateauing at a suboptimal threshold—a consequence of the exponential amplification mechanism indiscriminately boosting both the target signal and background noise. Conversely, the Off-ED configuration robustly sustains a high-fidelity SNR plateau at elevated $\Delta\kappa_{2}$ values, thereby circumventing the noise-amplification crisis. While the singularity provides giant sensitivity for infinitesimal inputs, its growth rate slackens for slightly larger perturbations, allowing the Off-ED regime (SNR $\propto \Delta\kappa_{2}$) to eventually overtake it and deliver superior signal clarity across a broader operational window.

To unveil the fundamental scaling laws governing this noise resilience, we inspect the SNR as a function of the noise amplitude $\delta$ under both operating conditions. At the strict ED point [Fig.~\ref{Fig5}(c)], the system follows an unfavorable power-law scaling characterized by $\text{SNR} \propto \delta^{-1/2}$. This square-root dependence is a double-edged sword: the very same geometric singularity that unlocks exponential sensitivity simultaneously renders the energy spectrum hypersensitive to stochastic noise, causing background fluctuations to scale as $\sqrt{\delta}$. Remarkable rectification is achieved by migrating to the Off-ED regime [Fig.~\ref{Fig5}(d)], where the system restores a clean linear scaling law, $\text{SNR} \propto \delta^{-1}$. Because a linear response ($\sim \delta$) suppresses weak noise far more efficiently than a square-root response ($\sim \sqrt{\delta}$) when $\delta \ll 1$, the Off-ED configuration achieves an SNR enhancement of several orders of magnitude in the deep weak-noise regime (e.g., at $\delta = 10^{-8}$), markedly surpassing the strict At-ED state. The excellent agreement between our numerical ensembles and the analytical scaling curves firmly establishes that deliberately detuning the system away from the strict ED limit is a powerful and general strategy for optimizing non-Hermitian sensors in realistic, noise-limited applications.

To uncover the microscopic mechanism behind this macroscopic stability and its limits, we provide an analytical estimation of the topological restructuring threshold $\kappa_{2}^{c}$ grounded in non-Bloch band theory~\cite{Yokomizo2019}. Within the generalised Brillouin zone (GBZ) framework~\cite{Yao2018, Yang2020}, the open-boundary eigenstates of chain A are characterised by a complex Bloch momentum $\beta_{A}$ satisfying the GBZ saddle-point condition $|\beta_{A}|^{2}= (t_{a}-\gamma_{a})/(t_{a}+\gamma_{a})$, so that the skin-localised wavefunction decays exponentially as $|\psi_{n}^{A}|\sim e^{-n/\lambda}$, where the non-Bloch decay length is given by $\lambda = 2 /\left[\ln{(t_{a}+\gamma_{a})/(t_{a}-\gamma_{a})}\right]$~\cite{Hatano1996, Zhang2020a}. Under the ED condition ($t_{a}=t_{b}$, $\gamma_{a}=-\gamma_{b}$), chains A and B share an identical GBZ, yielding a single decay length $\lambda$ that governs the composite system. The introduction of a reverse coupling $\kappa_{2}$ establishes a macroscopic nonreciprocal loop in real space: the backward leakage amplitude arriving at the opposite boundary scales as $e^{-N/\lambda}$, and the complete round-trip across the double-chain geometry, requiring propagation through both chains, introduces an additional factor $e^{-N/\lambda}$, giving a total exponential suppression of $e^{-2N/\lambda}$. Balancing this attenuation against the reverse coupling strength yields the topological restructuring condition $\kappa_{1}\kappa_{2}^{c} \sim e^{-2N/\lambda}$, or equivalently, $\kappa_{2}^{c} \sim \frac{1}{\kappa_{1}} \exp\!\left(-\frac{2N}{\lambda}\right)$. Given the representative parameters $t_{a}=t_{b}=1, \gamma_{a}=0.5, \gamma_b=0.5$, and $\kappa_{1}=1$, the formulation yields $\lambda = 2/\ln 3 \approx 1.82$. The resulting predictions for the critical thresholds, specifically, $\kappa_{2}^{c} \approx 0.11$ for $N=2$, $\kappa_{2}^{c} \approx 1.2\times10^{-2}$ for $N=4$, $\kappa_{2}^{c} \approx 1.4\times10^{-3}$ for $N=6$, and $\kappa_{2}^{c} \approx 1.5\times10^{-4}$ for $N=8$, exhibit quantitative agreement with the spike positions marked by the dashed vertical lines in Fig.~\ref{Fig4}(c), with deviations remaining within a factor of two across all system sizes. These residual deviations are attributable to finite-size corrections beyond the leading exponential term and to the approximation of replacing the discrete lattice sum by its continuum GBZ counterpart. Beyond the critical threshold. Beyond the critical threshold ($\kappa_{2} > \kappa_{2}^{c}$), the macroscopic loop transmission dominates over the boundary nonreciprocity, inducing a profound topological restructuring of the energy spectrum in which the open-boundary line arcs transition into complex loops. Consequently, the NHSE is entirely quenched, terminating the size-dependent exponential sensitivity enhancement. Ultimately, this interplay between boundary exponential decay and reverse coupling establishes the rigorous operational bounds of the Off-ED sensing paradigm.

Armed with this macro-topological understanding of the background bias boundary, we proceed to dissect the individual micro-contributions within the noise matrix. To identify which specific channels within the global noise matrix predominantly govern the sensing performance, we project $\boldsymbol{\Omega}_{2N \times 2N}$ onto the real-space block basis corresponding to the double-chain architecture
\begin{equation}\label{eq:noise_block}
   \boldsymbol{\Omega}_{2N \times 2N} =
       \begin{pmatrix}
       \boldsymbol{\Omega}_{AA} & \boldsymbol{\Omega}_{AB} \\
       \boldsymbol{\Omega}_{BA} & \boldsymbol{\Omega}_{BB}
\end{pmatrix},
\end{equation}
where each sub-block $\boldsymbol{\Omega}{ij}$ ($i, j \in \{A, B\}$) is an $N \times N$ matrix. Specifically, $\boldsymbol{\Omega}_{AA}$ ($\boldsymbol{\Omega}_{BB}$) represents the intralayer noise modulating chain A (B), while $\boldsymbol{\Omega}_{BA}$ and $\boldsymbol{\Omega}_{AB}$ denote the interlayer noise blocks modulating the forward ($B \rightarrow A$) and reverse ($A \rightarrow B$) coupling pathways, respectively. Individual matrix elements within these sub-blocks are systematically indexed as $\Omega_{\eta\nu}^{(i,j)}$ ($\eta,\nu \in \{A,B\}$ and $i,j \in \{1, \dots, N\}$).

\begin{figure}[t!]
\centering
\includegraphics[width=0.40\textwidth,keepaspectratio]{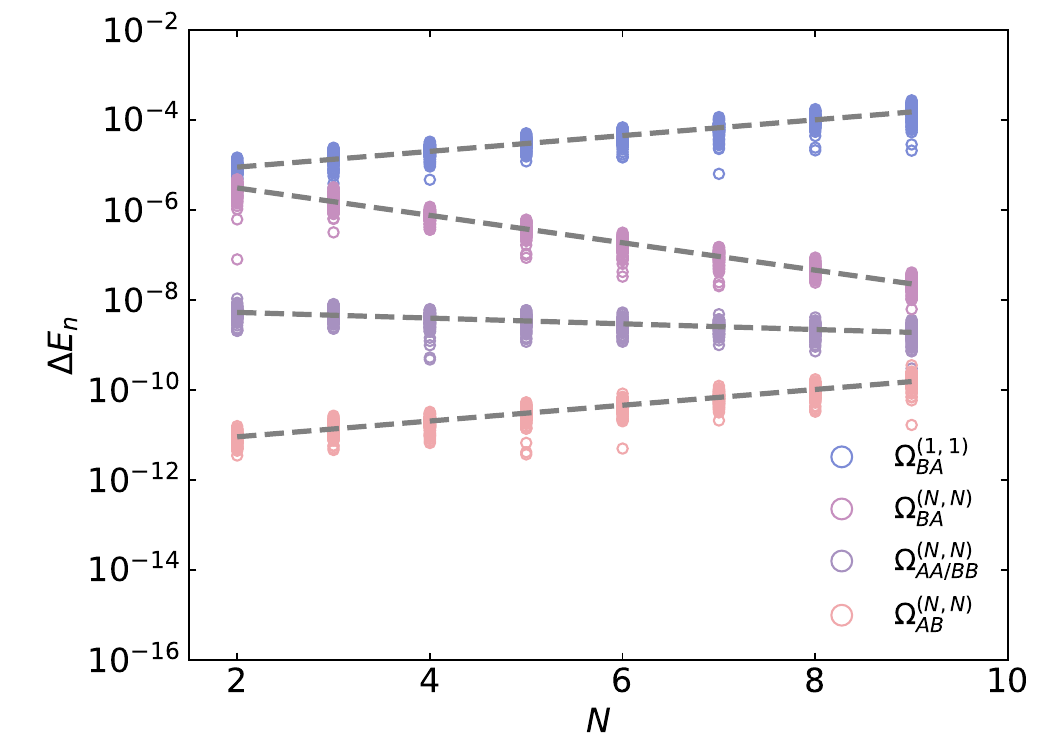}
\caption{Noise-induced eigenspectral splitting $\Delta E_n$ as a function of system size $N$ for four representative noise matrix elements in the Off-ED regime ($\kappa_2 = 10^{-6}$) at a fixed noise level $\delta = 10^{-8}$. Data points correspond to four channels listed in order of decreasing noise amplification strength: (i) the left-boundary interlayer element $\Omega^{(1,1)}_{BA}$ (blue circles), which maximally overlaps with the skin-localized wavefunctions and produces exponentially growing $\Delta E_n$; (ii) the right-boundary interlayer element $\Omega^{(N,N)}_{BA}$ (purple circles), which is spatially separated from the skin modes and yields exponentially decaying $\Delta E_n$; (iii) the intralayer diagonal element $\Omega^{(N,N)}_{AA/BB}$ (lavender circles), which produces a size-independent noise floor; and (iv) the reverse interlayer element $\Omega^{(N,N)}_{AB}$ (pink circles), which is subject to dual suppression from both wavefunction separation and coupling directionality, rendering it the most deeply suppressed channel. Dashed lines indicate the corresponding analytical scaling references for each individual component.}
\label{Fig6}
\end{figure}

To establish the theoretical foundation of our noise-matrix approach and its connection to the conventional non-Hermitian sensing framework, we derive an explicit bridge between the noise matrix $\boldsymbol{\Omega}$ and a generalized Petermann factor~\cite{Zhang2019a, Wiersig2020}. In a conventional single-mode EP sensor, a stochastic perturbation induces an eigenvalue fluctuation whose ensemble variance scales with the scalar Petermann factor. However, in the ED architecture investigated here, the system hosts a proliferation of independent Jordan blocks, each characterized by a distinct local dimension of $M=2$~\cite{Li2026a}. Although the dimension of any single block is independent of the system scale, the macroscopic cooperative phenomena emerging from the $M=2$ defective subspace spanning $N$ sites necessitate a multi-mode generalized Petermann matrix approach to rigorously assess the global noise sensitivity. This block is spanned by the right generalized eigenvectors $|u_{n,1}\rangle$, $|u_{n,0}\rangle$ and the left generalized eigenvectors $\langle v_{n,1}|$, $\langle v_{n,0}|$, which are bound by strict algebraic recurrence relations
\begin{align}
	(H_0 - E_n)|u_{n,1}\rangle = |u_{n,0}\rangle, (H_0 - E_n)|u_{n,0}\rangle = 0,\\
	(H_0^\dagger - E_n^*)|v_{n,1}\rangle = |v_{n,0}\rangle, (H_0^\dagger - E_n^*)|v_{n,0}\rangle = 0.
\end{align}

Under such extreme eigenspace aggregation, a scalar description fails to capture the collective multi-mode noise response driven by the entire Jordan block. Crucially, the conventional denominator $\lvert\langle v_0|u_0\rangle\rvert^2$~\cite{Zhang2019a} strictly vanishes during high-dimensional block degeneration, causing a mathematical breakdown. To address this challenge, we adopt a rigorous biorthogonal normalization condition $\langle v_{n,k} | u_{n,l} \rangle = \delta_{k, 1-l} \langle v_{n,0} | u_{n,1} \rangle$ (with $\langle v_{n,0} | u_{n,1} \rangle \neq 0$)~\cite{Seyranian2003}, and introduce a properly regularized matrix-valued generalized Petermann factor defined as
\begin{align}
	\mathbf{K}_{n}^{(\text{ED})} = \frac{\langle v_{n,k} | v_{n,k} \rangle \langle u_{n,l} | u_{n,l} \rangle}{|\langle v_{n,0} | u_{n,1} \rangle|^2}, \quad k,l \in \{0,1\},
\end{align}
where $n \in \{1, \dots, N\}$ encompasses the entire unperturbed spectral range. Owing to the macroscopic cooperative phenomena emerging from these $M=2$ defective subspaces, the global noise amplification profile is dictated by the total trace of the sub-block matrices, namely $\text{tr}(\mathbf{K}^{(\text{ED})}) = \sum_{n=1}^N \text{tr}(\mathbf{K}_n^{(\text{ED})})$. Remarkably, this total trace exhibits a strict linear scaling with the lattice size $N$ ($\text{tr}(\mathbf{K}^{(\text{ED})}) \propto N$)~\cite{Yang2020, Zhang2019a, Li2026a}.

To guarantee statistical-mechanical consistency alongside the self-consistency of numerical computations under OBCs, one must account for the real-space distribution profiles of the generalized eigenvectors. Due to the opposing surface-localization directions of chains A and B, the left eigenvector $\langle v_A |$ and the right eigenvector $| u_B \rangle$ accumulate at opposite boundaries. This pronounced spatial separation establishes the "Zero-Overlap Trap"~\cite{Zhang2019a, Yokomizo2019, Yang2020, Li2026a}, whereby the scaling behavior of the effective matrix elements is exponentially suppressed by the overlap integral, expressed as $I_{\text{overlap}} = |\sum_{n=1}^N \langle v_A | n \rangle \langle n | u_B \rangle| \propto N e^{-N/\lambda}$. 

Furthermore, directly evaluating the spectral variance via $\langle|\Delta E_n|^2\rangle \propto \sigma^2 \mathrm{tr}(\mathbf{K}^{(\mathrm{ED})}\cdot\mathbf{M}_{\Omega})$, with $\mathbf{M}_{\Omega}$ denoting the projection of the noise matrix onto the generalized basis, introduces an inherent causal mismatch and domain incompatibility at the strict ED limit. In this regime, the fractional-power spectral splitting physically originates from the non-diagonalizable nature of the coalesced eigenspace, rendering the conventional variance representation fundamentally inadequate. Specifically, because the non-diagonalizable Jordan-block structure produces a branch-cut singularity in the perturbative expansion of the secular equation, the eigenvalue response to a stochastic perturbation $\delta$ is intrinsically non-analytic: for each local Jordan block of dimension $M=2$, the eigenvalue shift exhibits a characteristic $\delta^{1/2}$ scaling behavior. This fractional-power response guarantees that the ensemble mean $\langle \Delta E_n \rangle \propto \delta^{1/2}$ remains uniquely consistent with the causal mechanism of physically observed spectral splitting. Conversely, while the conventional variance formulation $\langle |\Delta E_n|^2 \rangle$ accounts for higher-order statistical invariants, the resolvent operator $(E - H_0)^{-1}$ undergoes a non-analytic algebraic divergence at the ED singularity. Consequently, the resulting probability density function departs entirely from Gaussian symmetry, making it impossible to directly map these statistical features onto first-order observables. Characterizing noise amplification through the variance $\langle|\Delta E_n|^2\rangle$ overestimates the effective noise floor by conflating the algebraic divergence of the resolvent with the physical spectral deformation. The spectral splitting is more appropriately characterized by its ensemble mean, which directly maps onto the experimentally observable eigenvalue displacement.

To reconcile the mathematical self-contradiction between the variance scaling and the fractional splitting ($\Delta E_n \propto \delta^{1/2}$), the noise-induced spectral splitting must be updated to its regularized form for $M=2$, which seamlessly integrates a spatial instability truncation framework~\cite{Kato1966, Lau2018}
\begin{align}
	\langle\Delta E_n\rangle \propto \left[ \delta^2 \cdot \text{tr}\left(\mathbf{K}^{(\text{ED})} \cdot \mathbf{M}_\Omega\right) \cdot N^2 e^{-2N/\lambda} \right]^{1/4},
\end{align}
which yields the following explicit asymptotic scaling behavior
\begin{align}\label{Eq21}
	\langle\Delta E_n\rangle \propto \delta^{1/2} N^{1/2} \left[\text{tr}\left(\mathbf{K}^{(\text{ED})} \cdot \mathbf{M}_\Omega\right)\right]^{1/4} \cdot e^{-N/(2\lambda)}.
\end{align}

This regularization framework rigorously reconciles the algebraic singularity of $M=2$ Jordan blocks with their spatial configurations under OBCs. Due to the opposing surface-localization directions of chains A and B, the left and right generalized eigenvectors undergo a pronounced real-space separation, establishing the "Zero-Overlap Trap"~\cite{Zhang2019a, Yokomizo2019, Yang2020, Li2026a}. This spatial segregation exponentially attenuates the effective interlayer perturbative coupling via the overlap integral $I_{\text{overlap}} \propto N e^{-N/\lambda}$. As a result, with increasing lattice size $N$, the exponential decay factor $e^{-N/(2\lambda)}$ heavily dominates over the feeble algebraic growth of $N^{1/2}$ and the linear scaling of $\text{tr}(\mathbf{K}^{(\text{ED})})$, abruptly quenching the singular spectral splitting response down to the machine precision limit. This resolution disposes of the pre-existing mathematical self-contradiction, providing a physically viable, non-zero boundary condition constraint for high-sensitivity sensors operating at the precise ED limit. 

Meanwhile, the numerical instability of eigenvalue-splitting sensors in the large-$N$ limit under strict At-ED conditions underscores the vital need to pivot toward the Off-ED region. In this region, a deterministic background bias $\kappa_2 \neq 0$ lifts the $M=2$ Jordan blocks into isolated, non-degenerate eigenstates, thoroughly avoiding the precision degradation induced by the spatial zero-overlap trap. Consequently, under the lowest-order Jordan degeneracy ($M=2$), the physically observable spectral splitting under finite numerical precision is rigorously regularized as Eq.~\eqref{Eq21}. As the lattice size $N$ grows, the exponential decay factor $e^{-N/(2\lambda)}$ severely outpaces both the weak algebraic growth of $N^{1/2}$ and the strict linear scaling of $\text{tr}(\mathbf{K}^{(\text{ED})}) \propto N$, causing the fractional response to rapidly plummet toward the machine precision floor. This formula explicitly demonstrates that, within an intermediate system-size window where the algebraic prefactor $N^{1/2}\cdot\left[\mathrm{tr}\!\left(\mathbf{K}^{(\mathrm{ED})}\cdot \mathbf{M}_{\Omega}\right)\right]^{1/4}$ has not yet been overwhelmed by the exponential decay factor $e^{-N/(2\lambda)}$, the noise-induced spectral splitting scales as $\langle\Delta E_{n}\rangle\propto\delta^{1/2}$, which fundamentally dictates the undesirable scaling relation $\mathrm{SNR}\propto\delta^{-1/2}$ at the exact At-ED point~\cite{Zhang2019a}. It should be noted that as $N$ grows beyond a system-size-dependent crossover $N_{c}$, the exponential suppression $e^{-N/(2\lambda)}$ dominates, driving $\langle\Delta E_{n}\rangle$ toward the machine-precision floor; in this large-$N$ limit the At-ED sensor loses its spectral resolution entirely and the $\delta^{-1/2}$ scaling becomes empirically inaccessible. The numerical results in Fig.~\ref{Fig5}(c) are therefore obtained in the regime $N<N_{c}$, where the fractional-order scaling is still observable, and the comparison with the Off-ED $\delta^{-1}$ law is physically meaningful within this same window.

In sharp contrast, entering the Off-ED regime by introducing a finite deterministic background bias $\kappa_2 \neq 0$ establishes a macroscopic non-Hermitian closed loop via the forward ($\kappa_1$) and reverse ($\kappa_2$) couplings. This operation fundamentally reconfigures the local algebraic structure of the secular equation for each independent mode $n$, yields $\det(\lambda I - H_{\text{local},n}) = (\lambda - E_n)^2 - \kappa_1 \kappa_2 = 0$, and effectively eradicates the geometric singularity at the origin. Consequently, the consolidated subspace splits into $N$ pairs of isolated, non-degenerate eigenstates, lifted by a robust and well-defined bias gap $\Delta E_{\text{bias}} \propto (\kappa_1 \kappa_2)^{1/2}$. This splitting behavior is strictly dictated by the intrinsic topology of the local $M=2$ Jordan blocks, rather than being an artifact of an artificially engineered higher-order singularity. When the external random perturbation is weak ($\delta \ll \kappa_2$), a first-order Taylor expansion of the secular equation yields a noise correction that is strictly linear in $\delta$~\cite{Zhong2019}. This mechanism successfully mitigates scaling divergences and eradicates the pathological algebraic singularities inherent in macroscopic Jordan blocks, thereby restoring the highly coveted linear scaling relation $\text{SNR} \propto \delta^{-1}$~\cite{Landers2026}.

As illustrated in Fig.~\ref{Fig6}, the noise-induced eigenspectral splitting $\Delta E_n$ exhibits vastly distinct scaling behaviors across the same four representative noise channels in the moderately Off-ED regime (background bias $\kappa_2 = 10^{-6}$), ordered here by decreasing amplification strength. For the forward B $\rightarrow$ A configuration, which features left-directed skin localization: (i) the left-boundary interlayer element $\Omega_{BA}^{(1,1)}$ maximizes the spatial overlap between the stochastic perturbation and the skin-localized wavefunctions. Consequently, the resulting spectral splitting $\Delta E_n$ grows exponentially with the lattice size $N$, establishing $\Omega_{BA}^{(1,1)}$ as the single most pathological gateway for noise amplification. In sharp contrast, (ii) the forward interlayer element located at the opposite edge, $\Omega_{BA}^{(N,N)}$, experiences a twofold spatial separation from the skin-localized state profile, and its contribution to $\Delta E_n$ undergoes rapid exponential decay with increasing $N$, demonstrating that the system possesses a robust form of topological immunity against local fluctuations at the right boundary. Meanwhile, (iii) the intralayer diagonal noise elements representing intrinsic cavity fluctuations, exemplified by $\Omega_{AA/BB}^{(N,N)}$, produce a size-independent, near-constant $\Delta E_n$, behaving as a conventional background noise floor that remains unaffected by non-Hermitian scaling dynamics. Most remarkably, (iv) the right-boundary element of the reverse interlayer channel, $\Omega_{AB}^{(N,N)}$, is subject to the simultaneous suppression of both physical wavefunction separation and the unidirectional nature of the interlayer coupling. This dual-layer protection renders $\Omega_{AB}^{(N,N)}$ the most deeply suppressed noise component, locking its spectral response within an infinitesimal window that can be entirely marginalized in the macroscopic sensing dynamics.

From a mathematical perspective, it is of paramount importance to distinguish the macroscopic scaling behavior of the deterministic target signal $\Delta\kappa_2$ from the microscopic quenching of the localized random noise element $\Omega_{AB}^{(N,N)}$ sharing the identical reverse interlayer channel. Mechanistically, while any localized perturbation in the reverse channel undergoes severe exponential suppression at the microscopic scale due to the spatial zero-overlap trap ($I_{\text{overlap}} \propto N e^{-N/\lambda}$), the deterministic target signal $\Delta\kappa_2$ is enforced globally across all spatial interfaces or specified boundaries. This collective enforcement constructs a coherent macroscopic non-reciprocal loop that successfully reinstates the global non-Bloch boundary conditions. Such macroscopic phase coherence allows the signal to dictate a dramatic restructuring of the spectral arc into a closed complex loop, thereby fully unlocking the size-dependent exponential sensitivity driven by the NHSE. Conversely, the random fluctuations in $\Omega_{AB}^{(N,N)}$ remain spatially uncorrelated and strictly localized, rendering them incapable of establishing a coherent round-trip transport loop. As a result, the noise channel is permanently frozen within the infinitesimal window dictated by the zero-overlap trap, while the macroscopic signal channel robustly sustains its amplification pathway. This asymmetric regularization mechanism serves as the ultimate physical cornerstone for the high SNR observed within the Off-ED regime.

These microscopic insights unveil the broader physical architecture of the sensitivity-noise trade-off in non-Hermitian systems. Operating strictly at the exact ED limit, the completely degenerate high-dimensional Jordan block induces an unfavorable power-law scaling ($\text{SNR} \propto \delta^{-1/2}$) and a non-analytic signal response ($\text{SNR} \propto (\Delta\kappa_2)^{1/2}$), rendering the energy spectrum hypersensitive to stochastic environmental fluctuations. Because both the target signal and background noise undergo identical macroscopic amplification via the skin effect, the ultimate SNR is severely capped despite being size-independent. In stark contrast, migrating slightly into the off-ED regime breaks the pathological algebraic singularity while leaving the real-space directional pumping entirely intact. On one hand, a minute background bias $\kappa_2$ dismantles the fine-tuned Jordan block structure, collapsing the eigenspace similarity $\mathcal{C}$ below unity. This algebraic collapse successfully truncates the square-root noise amplification channel and restores a clean linear scaling relation ($\text{SNR} \propto \Delta\kappa_2$, $\text{SNR} \propto \delta^{-1}$), thereby profoundly depressing the noise floor. On the other hand, as long as $\kappa_2$ is kept well below the topological reconfiguration threshold ($\kappa_2 \ll \gamma_{a,b}$), the intra-chain asymmetric hopping ($t_{a,b} \pm \gamma_{a,b}$) perpetually sustains deep boundary localization. This ensures that the exponential sensitivity scaling powered by the NHSE is fully retained, while completely circumventing the singularity-induced noise inflation. Crucially, shifting to the Off-ED operating paradigm not only reinstates the linear scaling law ($\text{SNR} \propto \delta^{-1}$) but also eradicates the numerical instabilities stemming from the zero-overlap trap. The introduction of a finite, deterministic background bias $\kappa_2 \neq 0$ allows the macroscopic non-Hermitian closed loop to obviate the need for a fine-tuned, exact boundary isolation mechanism. This guarantees a robust reconstruction of sensing performance without requiring unphysically large lattice scales N that would otherwise submerge the singular derivative response into the machine-precision floor. This profound decoupling between algebraic fragility and topological skin amplification definitively underpins the Off-ED configuration as a self-consistent, optimal operation framework for dual-chain Hatano–Nelson sensors, providing a highly practical blueprint for targeted noise reduction in realistic physical environments. We further emphasize that the introduction of a finite background bias $\kappa_2$ must be carefully optimized to balance the recovery of linear scaling behavior against the dynamic range constraints inherent to practical detection platforms. Although a larger $\kappa_2$ can guarantee robust linearity (SNR $\propto \delta^{-1}$) by enforcing the zero-overlap trapping condition, it simultaneously generates a macroscopic baseline energy gap $\Delta E\propto \sqrt{\kappa_2}$. In practical implementations based on optical crystalline nodes or topological electrical networks, this substantial baseline offset demands sufficient dynamic range and high precision from the readout electronics, so as to resolve minute target variations $\Delta\kappa_2$ against a considerably larger DC background signal. Consequently, maintaining $\kappa_2$ within an intermediate regime — appreciably below the topological reconstruction threshold $\kappa_{2}^{c}$, yet still well above the minimum resolution limit of the device — constitutes the ideal engineering parameter choice for next-generation non-Hermitian sensors.

\section{Conclusion}
\label{sec:conclusion}

In summary, we have systematically investigated a non-Hermitian sensing architecture featuring ED in a double-chain Hatano–Nelson model. While operating strictly at the exact ED singularity enforces a hypersensitive yet noise-susceptible fractional response that caps the performance ceiling, migrating slightly into the Off-ED regime restores a clean linear scaling regime. This micro-biased paradigm successfully circumvents the singularity-induced elevation of the noise floor while fully preserving the macroscopic exponential sensitivity driven by the NHSE. Microscopic analysis of the noise matrix further reveals a highly asymmetric noise landscape: the left-boundary interlayer channel is pinpointed as the dominant vulnerability, producing exponentially growing spectral splitting that sets the practical noise ceiling. In sharp contrast, the right-boundary interlayer channel and, most profoundly, the reverse interlayer channel are subject to exponential suppression arising from the spatial zero-overlap trap, exhibiting a topological immunity that can be exploited as an additional degree of freedom for directional noise filtering. Together, these findings provide a concrete, channel-resolved blueprint for targeted noise mitigation in realistic physical environments.

This Off-ED sensing framework establishes a powerful design principle that successfully reconciles extreme sensitivity with robust noise immunity by exploiting the algebraic fragility of high-dimensional Jordan blocks. The theoretical insights established herein not only elucidate the spatial fluctuation dynamics in nonreciprocal lattices, but also point toward immediate experimental implementation. Owing to its structural generality, the proposed double-chain model and its disorder perturbation mechanics can be seamlessly mapped onto a variety of prominent platforms, including nonreciprocal topological circuit networks, acoustic metamaterial resonator arrays, and photonic crystal waveguide lattices~\cite{Zhang2021a, Song2020, Weidemann2020, Imhof2018}.

\section*{Acknowledgments}

This work is supported by the National Natural Science Foundation of China (11975175).

\section*{DATA AVAILABILITY STATEMENT}

All raw data corresponding to the findings in this paper are available from the authors upon
reasonable request.

\section*{Author declarations}

The authors have no conflicts to disclose.

\bibliography{references_main.bib}

@Article{Li2026a,
  author  = {Li, Zhen and Cai, Rundong and Wang, Xulong and Shimomura, Kenji and Lu, Congwei and Yang, Zhesen and Sato, Masatoshi and Ma, Guancong},
  journal = {Nat. Phys.},
  title   = {Exceptional deficiency of non-{H}ermitian systems},
  year    = {2026},
  issn    = {1745-2481},
  doi     = {10.1038/s41567-026-03259-7},
  refid   = {Li2026},
  url     = {https://doi.org/10.1038/s41567-026-03259-7},
}

@Article{Bid2026,
  author    = {Bid, Subhajyoti and Schomerus, Henning},
  journal   = {Phys. Rev. Res.},
  title     = {Exceptionally deficient topological square-root insulators},
  year      = {2026},
  month     = {Feb},
  pages     = {L012031},
  volume    = {8},
  doi       = {10.1103/svp4-ksxf},
  issue     = {1},
  numpages  = {6},
  publisher = {American Physical Society},
  url       = {https://link.aps.org/doi/10.1103/svp4-ksxf},
}

@Article{Bender1998,
  author    = {Bender, Carl M. and Boettcher, Stefan},
  journal   = {Phys. Rev. Lett.},
  title     = {Real Spectra in Non-{H}ermitian Hamiltonians Having $\mathcal{P}\mathcal{T}$ Symmetry},
  year      = {1998},
  month     = {Jun},
  pages     = {5243--5246},
  volume    = {80},
  doi       = {10.1103/PhysRevLett.80.5243},
  issue     = {24},
  numpages  = {0},
  publisher = {American Physical Society},
  url       = {https://link.aps.org/doi/10.1103/PhysRevLett.80.5243},
}

@Article{Ashida2020,
  author    = {Ashida, Yuto and Gong, Zongping and Ueda, Masahito},
  journal   = {Adv. Phys.},
  title     = {Non-{H}ermitian physics},
  year      = {2020},
  issn      = {0001-8732},
  month     = jul,
  number    = {3},
  pages     = {249--435},
  volume    = {69},
  comment   = {doi: 10.1080/00018732.2021.1876991},
  doi       = {10.1080/00018732.2021.1876991},
  publisher = {Taylor & Francis},
  url       = {https://doi.org/10.1080/00018732.2021.1876991},
}

@Article{Okuma2023,
  author    = {Okuma, Nobuyuki and Sato, Masatoshi},
  journal   = {Annu. Rev. Condens. Matter Phys.},
  title     = {Non-{H}ermitian Topological Phenomena: A Review},
  year      = {2023},
  issn      = {1947-5462},
  number    = {Volume 14, 2023},
  pages     = {83-107},
  volume    = {14},
  doi       = {https://doi.org/10.1146/annurev-conmatphys-040521-033133},
  publisher = {Annual Reviews},
  type      = {Journal Article},
  url       = {https://www.annualreviews.org/content/journals/10.1146/annurev-conmatphys-040521-033133},
}

@Article{Kawabata2019,
  author    = {Kawabata, Kohei and Shiozaki, Ken and Ueda, Masahito and Sato, Masatoshi},
  journal   = {Phys. Rev. X},
  title     = {Symmetry and Topology in Non-{H}ermitian Physics},
  year      = {2019},
  month     = oct,
  number    = {4},
  pages     = {041015},
  volume    = {9},
  doi       = {10.1103/PhysRevX.9.041015},
  publisher = {American Physical Society},
  refid     = {10.1103/PhysRevX.9.041015},
  url       = {https://link.aps.org/doi/10.1103/PhysRevX.9.041015},
}

@Article{ElGanainy2018,
  author  = {El-Ganainy, Ramy and Makris, Konstantinos G. and Khajavikhan, Mercedeh and Musslimani, Ziad H. and Rotter, Stefan and Christodoulides, Demetrios N.},
  journal = {Nat. Phys.},
  title   = {Non-{H}ermitian physics and PT symmetry},
  year    = {2018},
  issn    = {1745-2481},
  number  = {1},
  pages   = {11--19},
  volume  = {14},
  doi     = {10.1038/nphys4323},
  refid   = {El-Ganainy2018},
  url     = {https://doi.org/10.1038/nphys4323},
}

@Article{Gong2018,
  author    = {Gong, Zongping and Ashida, Yuto and Kawabata, Kohei and Takasan, Kazuaki and Higashikawa, Sho and Ueda, Masahito},
  journal   = {Phys. Rev. X},
  title     = {Topological Phases of Non-{H}ermitian Systems},
  year      = {2018},
  month     = sep,
  number    = {3},
  pages     = {031079},
  volume    = {8},
  doi       = {10.1103/PhysRevX.8.031079},
  publisher = {American Physical Society},
  refid     = {10.1103/PhysRevX.8.031079},
  url       = {https://link.aps.org/doi/10.1103/PhysRevX.8.031079},
}

@Article{Heiss2012,
  author    = {Heiss, W D},
  journal   = {J. Phys. A Math. Theor.},
  title     = {The physics of exceptional points},
  year      = {2012},
  month     = {oct},
  number    = {44},
  pages     = {444016},
  volume    = {45},
  doi       = {10.1088/1751-8113/45/44/444016},
  publisher = {IOP Publishing},
  url       = {https://dx.doi.org/10.1088/1751-8113/45/44/444016},
}

@Article{CanosValero2025,
  author    = {Can\'os Valero, Adri\`a and Sztranyovszky, Zoltan and Muljarov, Egor A. and Bogdanov, Andrey and Weiss, Thomas},
  journal   = {Phys. Rev. Lett.},
  title     = {Exceptional Bound States in the Continuum},
  year      = {2025},
  month     = {Mar},
  pages     = {103802},
  volume    = {134},
  doi       = {10.1103/PhysRevLett.134.103802},
  issue     = {10},
  numpages  = {7},
  publisher = {American Physical Society},
  url       = {https://link.aps.org/doi/10.1103/PhysRevLett.134.103802},
}

@Article{Bergholtz2021,
  author    = {Bergholtz, Emil J. and Budich, Jan Carl and Kunst, Flore K.},
  journal   = {Rev. Mod. Phys.},
  title     = {Exceptional topology of non-{H}ermitian systems},
  year      = {2021},
  month     = feb,
  number    = {1},
  pages     = {015005},
  volume    = {93},
  doi       = {10.1103/RevModPhys.93.015005},
  publisher = {American Physical Society},
  refid     = {10.1103/RevModPhys.93.015005},
  url       = {https://link.aps.org/doi/10.1103/RevModPhys.93.015005},
}

@Article{Mandal2021,
  author    = {Mandal, Ipsita and Bergholtz, Emil J.},
  journal   = {Phys. Rev. Lett.},
  title     = {Symmetry and Higher-Order Exceptional Points},
  year      = {2021},
  month     = {Oct},
  pages     = {186601},
  volume    = {127},
  doi       = {10.1103/PhysRevLett.127.186601},
  issue     = {18},
  numpages  = {6},
  publisher = {American Physical Society},
  url       = {https://link.aps.org/doi/10.1103/PhysRevLett.127.186601},
}

@Article{Tsuneya2026,
	title={{Hopf exceptional points}},
	author={Tsuneya Yoshida and Emil J. Bergholtz and Tomáš Bzdušek},
	journal={SciPost Phys.},
	volume={20},
	pages={001},
	year={2026},
	publisher={SciPost},
	doi={10.21468/SciPostPhys.20.1.001},
	url={https://scipost.org/10.21468/SciPostPhys.20.1.001},
}

@Article{Lin2025,
  author  = {Lin, Jhen-Dong and Kuo, Po-Chen and Lambert, Neill and Miranowicz, Adam and Nori, Franco and Chen, Yueh-Nan},
  journal = {Nat. Commun.},
  title   = {Non-{M}arkovian quantum exceptional points},
  year    = {2025},
  issn    = {2041-1723},
  number  = {1},
  pages   = {1289},
  volume  = {16},
  doi     = {10.1038/s41467-025-56242-w},
  refid   = {Lin2025},
  url     = {https://doi.org/10.1038/s41467-025-56242-w},
}

@Article{Delplace2021,
  author    = {Delplace, Pierre and Yoshida, Tsuneya and Hatsugai, Yasuhiro},
  journal   = {Phys. Rev. Lett.},
  title     = {Symmetry-Protected Multifold Exceptional Points and Their Topological Characterization},
  year      = {2021},
  month     = {Oct},
  pages     = {186602},
  volume    = {127},
  doi       = {10.1103/PhysRevLett.127.186602},
  issue     = {18},
  numpages  = {6},
  publisher = {American Physical Society},
  url       = {https://link.aps.org/doi/10.1103/PhysRevLett.127.186602},
}

@Article{Xiao2021,
  author    = {Xiao, Lei and Deng, Tianshu and Wang, Kunkun and Wang, Zhong and Yi, Wei and Xue, Peng},
  journal   = {Phys. Rev. Lett.},
  title     = {Observation of Non-{B}loch Parity-Time Symmetry and Exceptional Points},
  year      = {2021},
  month     = {Jun},
  pages     = {230402},
  volume    = {126},
  doi       = {10.1103/PhysRevLett.126.230402},
  issue     = {23},
  numpages  = {6},
  publisher = {American Physical Society},
  url       = {https://link.aps.org/doi/10.1103/PhysRevLett.126.230402},
}

@Article{Kononchuk2022,
  author  = {Kononchuk, Rodion and Cai, Jizhe and Ellis, Fred and Thevamaran, Ramathasan and Kottos, Tsampikos},
  journal = {Nature},
  title   = {Exceptional-point-based accelerometers with enhanced signal-to-noise ratio},
  year    = {2022},
  issn    = {1476-4687},
  number  = {7920},
  pages   = {697--702},
  volume  = {607},
  doi     = {10.1038/s41586-022-04904-w},
  refid   = {Kononchuk2022},
  url     = {https://doi.org/10.1038/s41586-022-04904-w},
}

@Article{Kawabata2019a,
  author    = {Kawabata, Kohei and Bessho, Takumi and Sato, Masatoshi},
  journal   = {Phys. Rev. Lett.},
  title     = {Classification of Exceptional Points and Non-{H}ermitian Topological Semimetals},
  year      = {2019},
  month     = {Aug},
  pages     = {066405},
  volume    = {123},
  doi       = {10.1103/PhysRevLett.123.066405},
  issue     = {6},
  numpages  = {7},
  publisher = {American Physical Society},
  url       = {https://link.aps.org/doi/10.1103/PhysRevLett.123.066405},
}

@Article{Ding2016,
  author    = {Ding, Kun and Ma, Guancong and Xiao, Meng and Zhang, Z. Q. and Chan, C. T.},
  journal   = {Phys. Rev. X},
  title     = {Emergence, Coalescence, and Topological Properties of Multiple Exceptional Points and Their Experimental Realization},
  year      = {2016},
  month     = {Apr},
  pages     = {021007},
  volume    = {6},
  doi       = {10.1103/PhysRevX.6.021007},
  issue     = {2},
  numpages  = {13},
  publisher = {American Physical Society},
  url       = {https://link.aps.org/doi/10.1103/PhysRevX.6.021007},
}

@Article{MartinezAlvarez2018,
  author    = {Martinez Alvarez, V. M. and Barrios Vargas, J. E. and Foa Torres, L. E. F.},
  journal   = {Phys. Rev. B},
  title     = {Non-{H}ermitian robust edge states in one dimension: Anomalous localization and eigenspace condensation at exceptional points},
  year      = {2018},
  month     = {Mar},
  pages     = {121401(R)},
  volume    = {97},
  doi       = {10.1103/PhysRevB.97.121401},
  issue     = {12},
  numpages  = {6},
  publisher = {American Physical Society},
  url       = {https://link.aps.org/doi/10.1103/PhysRevB.97.121401},
}

@Article{Naghiloo2019,
  author  = {Naghiloo, M. and Abbasi, M. and Joglekar, Yogesh N. and Murch, K. W.},
  journal = {Nat. Phys.},
  title   = {Quantum state tomography across the exceptional point in a single dissipative qubit},
  year    = {2019},
  issn    = {1745-2481},
  number  = {12},
  pages   = {1232--1236},
  volume  = {15},
  doi     = {10.1038/s41567-019-0652-z},
  refid   = {Naghiloo2019},
  url     = {https://doi.org/10.1038/s41567-019-0652-z},
}

@Article{Ding2022,
  author  = {Ding, Kun and Fang, Chen and Ma, Guancong},
  journal = {Nat. Rev. Phys.},
  title   = {Non-{H}ermitian topology and exceptional-point geometries},
  year    = {2022},
  issn    = {2522-5820},
  number  = {12},
  pages   = {745--760},
  volume  = {4},
  doi     = {10.1038/s42254-022-00516-5},
  refid   = {Ding2022},
  url     = {https://doi.org/10.1038/s42254-022-00516-5},
}

@Article{Okuma2020,
  author    = {Okuma, Nobuyuki and Kawabata, Kohei and Shiozaki, Ken and Sato, Masatoshi},
  journal   = {Phys. Rev. Lett.},
  title     = {Topological Origin of Non-{H}ermitian Skin Effects},
  year      = {2020},
  month     = {Feb},
  pages     = {086801},
  volume    = {124},
  doi       = {10.1103/PhysRevLett.124.086801},
  issue     = {8},
  numpages  = {7},
  publisher = {American Physical Society},
  url       = {https://link.aps.org/doi/10.1103/PhysRevLett.124.086801},
}

@Article{Song2019,
  author    = {Song, Fei and Yao, Shunyu and Wang, Zhong},
  journal   = {Phys. Rev. Lett.},
  title     = {Non-{H}ermitian Skin Effect and Chiral Damping in Open Quantum Systems},
  year      = {2019},
  month     = {Oct},
  pages     = {170401},
  volume    = {123},
  doi       = {10.1103/PhysRevLett.123.170401},
  issue     = {17},
  numpages  = {8},
  publisher = {American Physical Society},
  url       = {https://link.aps.org/doi/10.1103/PhysRevLett.123.170401},
}

@Article{Yokomizo2021,
  author    = {Yokomizo, Kazuki and Murakami, Shuichi},
  journal   = {Phys. Rev. B},
  title     = {Scaling rule for the critical non-{H}ermitian skin effect},
  year      = {2021},
  month     = {Oct},
  pages     = {165117},
  volume    = {104},
  doi       = {10.1103/PhysRevB.104.165117},
  issue     = {16},
  numpages  = {7},
  publisher = {American Physical Society},
  url       = {https://link.aps.org/doi/10.1103/PhysRevB.104.165117},
}

@Article{Li2020a,
  author  = {Li, Linhu and Lee, Ching Hua and Mu, Sen and Gong, Jiangbin},
  journal = {Nat. Commun.},
  title   = {Critical non-{H}ermitian skin effect},
  year    = {2020},
  issn    = {2041-1723},
  number  = {1},
  pages   = {5491},
  volume  = {11},
  doi     = {10.1038/s41467-020-18917-4},
  refid   = {Li2020},
  url     = {https://doi.org/10.1038/s41467-020-18917-4},
}

@Article{Jiang2019,
  author    = {Jiang, Hui and Lang, Li-Jun and Yang, Chao and Zhu, Shi-Liang and Chen, Shu},
  journal   = {Phys. Rev. B},
  title     = {Interplay of non-{H}ermitian skin effects and {A}nderson localization in nonreciprocal quasiperiodic lattices},
  year      = {2019},
  month     = {Aug},
  pages     = {054301},
  volume    = {100},
  doi       = {10.1103/PhysRevB.100.054301},
  issue     = {5},
  numpages  = {8},
  publisher = {American Physical Society},
  url       = {https://link.aps.org/doi/10.1103/PhysRevB.100.054301},
}

@Article{Zhang2021a,
  author  = {Zhang, Li and Yang, Yihao and Ge, Yong and Guan, Yi-Jun and Chen, Qiaolu and Yan, Qinghui and Chen, Fujia and Xi, Rui and Li, Yuanzhen and Jia, Ding and Yuan, Shou-Qi and Sun, Hong-Xiang and Chen, Hongsheng and Zhang, Baile},
  journal = {Nat. Commun.},
  title   = {Acoustic non-{H}ermitian skin effect from twisted winding topology},
  year    = {2021},
  issn    = {2041-1723},
  number  = {1},
  pages   = {6297},
  volume  = {12},
  doi     = {10.1038/s41467-021-26619-8},
  refid   = {Zhang2021},
  url     = {https://doi.org/10.1038/s41467-021-26619-8},
}

@Article{Zhang2021b,
  author  = {Zhang, Xiujuan and Tian, Yuan and Jiang, Jian-Hua and Lu, Ming-Hui and Chen, Yan-Feng},
  journal = {Nat. Commun.},
  title   = {Observation of higher-order non-{H}ermitian skin effect},
  year    = {2021},
  issn    = {2041-1723},
  number  = {1},
  pages   = {5377},
  volume  = {12},
  doi     = {10.1038/s41467-021-25716-y},
  refid   = {Zhang2021},
  url     = {https://doi.org/10.1038/s41467-021-25716-y},
}

@Article{Kawabata2020a,
  author    = {Kawabata, Kohei and Sato, Masatoshi and Shiozaki, Ken},
  journal   = {Phys. Rev. B},
  title     = {Higher-order non-{H}ermitian skin effect},
  year      = {2020},
  month     = {Nov},
  pages     = {205118},
  volume    = {102},
  doi       = {10.1103/PhysRevB.102.205118},
  issue     = {20},
  numpages  = {16},
  publisher = {American Physical Society},
  url       = {https://link.aps.org/doi/10.1103/PhysRevB.102.205118},
}

@Article{Lin2023,
  author  = {Lin, Rijia and Tai, Tommy and Li, Linhu and Lee, Ching Hua},
  journal = {Front. Phys.},
  title   = {Topological non-{H}ermitian skin effect},
  year    = {2023},
  issn    = {2095-0470},
  number  = {5},
  pages   = {53605},
  volume  = {18},
  doi     = {10.1007/s11467-023-1309-z},
  refid   = {Lin2023},
  url     = {https://doi.org/10.1007/s11467-023-1309-z},
}

@Article{Liang2022,
  author    = {Liang, Qian and Xie, Dizhou and Dong, Zhaoli and Li, Haowei and Li, Hang and Gadway, Bryce and Yi, Wei and Yan, Bo},
  journal   = {Phys. Rev. Lett.},
  title     = {Dynamic Signatures of Non-{H}ermitian Skin Effect and Topology in Ultracold Atoms},
  year      = {2022},
  month     = {Aug},
  pages     = {070401},
  volume    = {129},
  doi       = {10.1103/PhysRevLett.129.070401},
  issue     = {7},
  numpages  = {6},
  publisher = {American Physical Society},
  url       = {https://link.aps.org/doi/10.1103/PhysRevLett.129.070401},
}

@Article{Kawabata2023,
  author    = {Kawabata, Kohei and Numasawa, Tokiro and Ryu, Shinsei},
  journal   = {Phys. Rev. X},
  title     = {Entanglement Phase Transition Induced by the Non-{H}ermitian Skin Effect},
  year      = {2023},
  month     = {Apr},
  pages     = {021007},
  volume    = {13},
  doi       = {10.1103/PhysRevX.13.021007},
  issue     = {2},
  numpages  = {26},
  publisher = {American Physical Society},
  url       = {https://link.aps.org/doi/10.1103/PhysRevX.13.021007},
}

@Article{Guo2021b,
  author    = {Guo, Cui-Xian and Liu, Chun-Hui and Zhao, Xiao-Ming and Liu, Yanxia and Chen, Shu},
  journal   = {Phys. Rev. Lett.},
  title     = {Exact Solution of Non-{H}ermitian Systems with Generalized Boundary Conditions: Size-Dependent Boundary Effect and Fragility of the Skin Effect},
  year      = {2021},
  month     = {Sep},
  pages     = {116801},
  volume    = {127},
  doi       = {10.1103/PhysRevLett.127.116801},
  issue     = {11},
  numpages  = {6},
  publisher = {American Physical Society},
  url       = {https://link.aps.org/doi/10.1103/PhysRevLett.127.116801},
}

@Article{Okugawa2020,
  author    = {Okugawa, Ryo and Takahashi, Ryo and Yokomizo, Kazuki},
  journal   = {Phys. Rev. B},
  title     = {Second-order topological non-{H}ermitian skin effects},
  year      = {2020},
  month     = {Dec},
  pages     = {241202},
  volume    = {102},
  doi       = {10.1103/PhysRevB.102.241202},
  issue     = {24},
  numpages  = {6},
  publisher = {American Physical Society},
  url       = {https://link.aps.org/doi/10.1103/PhysRevB.102.241202},
}

@Article{Phan2026,
  author    = {Phan, Huyen Thanh and Wakabayashi, Katsunori},
  journal   = {Phys. Rev. B},
  title     = {Non-{H}ermitian corner skin effect in a two-dimensional photonic crystal},
  year      = {2026},
  month     = {May},
  pages     = {205403},
  volume    = {113},
  doi       = {10.1103/gvcy-vsvd},
  issue     = {20},
  numpages  = {9},
  publisher = {American Physical Society},
  url       = {https://link.aps.org/doi/10.1103/gvcy-vsvd},
}

@Article{Claes2021,
  author    = {Claes, Jahan and Hughes, Taylor L.},
  journal   = {Phys. Rev. B},
  title     = {Skin effect and winding number in disordered non-{H}ermitian systems},
  year      = {2021},
  month     = {Apr},
  pages     = {L140201},
  volume    = {103},
  doi       = {10.1103/PhysRevB.103.L140201},
  issue     = {14},
  numpages  = {7},
  publisher = {American Physical Society},
  url       = {https://link.aps.org/doi/10.1103/PhysRevB.103.L140201},
}

@Article{Gao2026,
  author    = {Gao, Nankun and Zhang, Xiujuan and Lu, Ming-Hui and Chen, Yan-Feng},
  journal   = {Phys. Rev. B},
  title     = {Multiwavepacket dynamics of non-{H}ermitian skin effects},
  year      = {2026},
  month     = {Mar},
  pages     = {094315},
  volume    = {113},
  doi       = {10.1103/2n6m-19ny},
  issue     = {9},
  numpages  = {10},
  publisher = {American Physical Society},
  url       = {https://link.aps.org/doi/10.1103/2n6m-19ny},
}

@Article{Rangi2025,
  author    = {Rangi, Chakradhar and Moreno, Juana and Tam, Ka-Ming},
  journal   = {Phys. Rev. B},
  title     = {Interplay of non-{H}ermitian skin effect and electronic correlations in the non-{H}ermitian {H}ubbard model via real-space dynamical mean-field theory},
  year      = {2025},
  month     = {Dec},
  pages     = {245137},
  volume    = {112},
  doi       = {10.1103/3nvj-g539},
  issue     = {24},
  numpages  = {8},
  publisher = {American Physical Society},
  url       = {https://link.aps.org/doi/10.1103/3nvj-g539},
}

@Article{Qin2025,
  author  = {Qin, Yi and Ang, Yee Sin and Lee, Ching Hua and Li, Linhu},
  journal = {Commun. Phys.},
  title   = {Many-body critical non-{H}ermitian skin effect},
  year    = {2025},
  issn    = {2399-3650},
  number  = {1},
  pages   = {16},
  volume  = {9},
  doi     = {10.1038/s42005-025-02448-9},
  refid   = {Qin2025},
  url     = {https://doi.org/10.1038/s42005-025-02448-9},
}

@Article{Cai2026,
  author  = {Cai, Xiaoming},
  journal = {Commun. Phys.},
  title   = {Non-{H}ermitian skin effect without point-gap topology in 2D quasicrystals},
  year    = {2026},
  issn    = {2399-3650},
  number  = {1},
  pages   = {61},
  volume  = {9},
  doi     = {10.1038/s42005-026-02496-9},
  refid   = {Cai2026},
  url     = {https://doi.org/10.1038/s42005-026-02496-9},
}

@Article{Shu2025,
  author    = {Shu, Chang and Zhang, Kai and Sun, Kai},
  journal   = {Phys. Rev. B},
  title     = {Ultraspectral sensitivity and nonlocal bound states in algebraic non-{H}ermitian skin effect},
  year      = {2025},
  month     = {Dec},
  pages     = {235152},
  volume    = {112},
  doi       = {10.1103/3927-n25r},
  issue     = {23},
  numpages  = {8},
  publisher = {American Physical Society},
  url       = {https://link.aps.org/doi/10.1103/3927-n25r},
}

@Article{Gliozzi2026,
  author    = {Gliozzi, Jacopo and Balducci, Federico and Hughes, Taylor L. and De Tomasi, Giuseppe},
  journal   = {Phys. Rev. B},
  title     = {Non-{H}ermitian multipole skin effects challenge localization},
  year      = {2026},
  month     = {Mar},
  pages     = {L100203},
  volume    = {113},
  doi       = {10.1103/c1jr-zqxg},
  issue     = {10},
  numpages  = {7},
  publisher = {American Physical Society},
  url       = {https://link.aps.org/doi/10.1103/c1jr-zqxg},
}

@Article{He2026,
  author    = {He, Peng and Lin, Sen},
  journal   = {Phys. Rev. B},
  title     = {Identifying the non-{H}ermitian skin effect in many-body dynamics by graphs},
  year      = {2026},
  month     = {Apr},
  pages     = {144302},
  volume    = {113},
  doi       = {10.1103/lxhl-wq7l},
  issue     = {14},
  numpages  = {9},
  publisher = {American Physical Society},
  url       = {https://link.aps.org/doi/10.1103/lxhl-wq7l},
}

@Article{Yang2026,
  author   = {Yang, Mengjie and Lee, Ching Hua},
  journal  = {Adv. Sci.},
  title    = {Beyond the Non-{H}ermitian Skin Effect: Scaling-Controlled Topology from Exceptional-Bound Bands},
  year     = {2026},
  number   = {20},
  pages    = {e23989},
  volume   = {13},
  doi      = {https://doi.org/10.1002/advs.202523989},
  url      = {https://advanced.onlinelibrary.wiley.com/doi/abs/10.1002/advs.202523989},
}

@Article{Li2026b,
  author    = {Li, Yang and Cai, Zhao-Fan and Liu, Tao and Nori, Franco},
  journal   = {Phys. Rev. B},
  title     = {Dissipation and interaction-controlled non-{H}ermitian skin effects},
  year      = {2026},
  month     = {Jan},
  pages     = {035444},
  volume    = {113},
  doi       = {10.1103/qrh6-vx64},
  issue     = {3},
  numpages  = {14},
  publisher = {American Physical Society},
  url       = {https://link.aps.org/doi/10.1103/qrh6-vx64},
}

@Article{Yang2026a,
  author    = {Yang, Xiaosen and Feng, Yaru and Wahab, Abdul and Geng, Hao},
  journal   = {Phys. Rev. A},
  title     = {Non-{H}ermitian second-order topological phases and bipolar skin effect in photonic kagome crystals},
  year      = {2026},
  month     = {Feb},
  pages     = {023506},
  volume    = {113},
  doi       = {10.1103/s26b-8bdl},
  issue     = {2},
  numpages  = {8},
  publisher = {American Physical Society},
  url       = {https://link.aps.org/doi/10.1103/s26b-8bdl},
}

@Article{Li2026c,
  author    = {Li, Jia-Rui and Jiang, Cui and Du, Kai and Zhang, Lian-Lian and Gong, Wei-Jiang},
  journal   = {Phys. Rev. B},
  title     = {Non-{H}ermitian skin effect and {M}ajorana signature in a {K}itaev chain with unidirectional hopping disorder},
  year      = {2026},
  month     = {Feb},
  pages     = {075419},
  volume    = {113},
  doi       = {10.1103/wxgq-vwk5},
  issue     = {7},
  numpages  = {16},
  publisher = {American Physical Society},
  url       = {https://link.aps.org/doi/10.1103/wxgq-vwk5},
}

@Article{Hu2025a,
  author    = {Hu, Yu-Min and Wang, Zijian and Lian, Biao and Wang, Zhong},
  journal   = {Phys. Rev. Lett.},
  title     = {Many-Body Non-{H}ermitian Skin Effect with Exact Steady States in the Dissipative Quantum Link Model},
  year      = {2025},
  month     = {Dec},
  pages     = {260401},
  volume    = {135},
  doi       = {10.1103/wztw-l8wg},
  issue     = {26},
  numpages  = {9},
  publisher = {American Physical Society},
  url       = {https://link.aps.org/doi/10.1103/wztw-l8wg},
}

@Article{Kunst2018,
  author    = {Kunst, Flore K. and Edvardsson, Elisabet and Budich, Jan Carl and Bergholtz, Emil J.},
  journal   = {Phys. Rev. Lett.},
  title     = {Biorthogonal Bulk-Boundary Correspondence in Non-{H}ermitian Systems},
  year      = {2018},
  month     = {Jul},
  pages     = {026808},
  volume    = {121},
  doi       = {10.1103/PhysRevLett.121.026808},
  issue     = {2},
  numpages  = {6},
  publisher = {American Physical Society},
  url       = {https://link.aps.org/doi/10.1103/PhysRevLett.121.026808},
}

@Article{Xiao2020,
  author  = {Xiao, Lei and Deng, Tianshu and Wang, Kunkun and Zhu, Gaoyan and Wang, Zhong and Yi, Wei and Xue, Peng},
  journal = {Nat. Phys.},
  title   = {Non-{H}ermitian bulk-boundary correspondence in quantum dynamics},
  year    = {2020},
  issn    = {1745-2481},
  number  = {7},
  pages   = {761--766},
  volume  = {16},
  doi     = {10.1038/s41567-020-0836-6},
  refid   = {Xiao2020},
  url     = {https://doi.org/10.1038/s41567-020-0836-6},
}

@Article{Yang2020,
  author    = {Yang, Zhesen and Zhang, Kai and Fang, Chen and Hu, Jiangping},
  journal   = {Phys. Rev. Lett.},
  title     = {Non-{H}ermitian Bulk-Boundary Correspondence and Auxiliary Generalized Brillouin Zone Theory},
  year      = {2020},
  month     = {Nov},
  pages     = {226402},
  volume    = {125},
  doi       = {10.1103/PhysRevLett.125.226402},
  issue     = {22},
  numpages  = {6},
  publisher = {American Physical Society},
  url       = {https://link.aps.org/doi/10.1103/PhysRevLett.125.226402},
}

@Article{Nakamura2024,
  author    = {Nakamura, Daichi and Bessho, Takumi and Sato, Masatoshi},
  journal   = {Phys. Rev. Lett.},
  title     = {Bulk-Boundary Correspondence in Point-Gap Topological Phases},
  year      = {2024},
  month     = {Mar},
  pages     = {136401},
  volume    = {132},
  doi       = {10.1103/PhysRevLett.132.136401},
  issue     = {13},
  numpages  = {10},
  publisher = {American Physical Society},
  url       = {https://link.aps.org/doi/10.1103/PhysRevLett.132.136401},
}

@Article{Zhang2026,
  author    = {Zhang, Xudong and Sun, Zhaoyu and Guo, Bin},
  journal   = {Phys. Rev. B},
  title     = {Unified topological bulk-entanglement correspondence for the polarization in non-{H}ermitian systems},
  year      = {2026},
  month     = {Mar},
  pages     = {L100401},
  volume    = {113},
  doi       = {10.1103/4pnb-mqmr},
  issue     = {10},
  numpages  = {8},
  publisher = {American Physical Society},
  url       = {https://link.aps.org/doi/10.1103/4pnb-mqmr},
}

@Article{Parasar2025,
  author    = {Parasar, Bhandaru Phani and Gefen, Yuval and Shenoy, Vijay B.},
  journal   = {Phys. Rev. Lett.},
  title     = {Bulk-Boundary Correspondence of Fractonic Field Theories},
  year      = {2025},
  month     = {Jun},
  pages     = {236601},
  volume    = {134},
  doi       = {10.1103/ggls-zhl8},
  issue     = {23},
  numpages  = {7},
  publisher = {American Physical Society},
  url       = {https://link.aps.org/doi/10.1103/ggls-zhl8},
}

@Article{Hu2025b,
  author    = {Hu, Kai-Xin and Luo, Chen and Zhang, Jie and Liu, Shutian and Cui, Wen-Xue and Cao, Ji and Zhang, Shou and Wang, Hong-Fu},
  journal   = {Phys. Rev. A},
  title     = {Bulk-boundary correspondence in non-{H}ermitian class-${\mathrm{D}}^{\ifmmode\dagger\else\textdagger\fi{}}$ topological systems},
  year      = {2025},
  month     = {Nov},
  pages     = {052209},
  volume    = {112},
  doi       = {10.1103/q3db-9zkj},
  issue     = {5},
  numpages  = {11},
  publisher = {American Physical Society},
  url       = {https://link.aps.org/doi/10.1103/q3db-9zkj},
}

@Article{Peng2025,
  author    = {Peng, Mian and Wei, Qiang and He, Ai-Lei and Huang, Zeheng and Deng, Weiyin and Liu, Zhengyou and Chen, Gang},
  journal   = {Phys. Rev. B},
  title     = {Multifold bulk-boundary correspondence in a gyromagnetic photonic crystal},
  year      = {2025},
  month     = {Dec},
  pages     = {245411},
  volume    = {112},
  doi       = {10.1103/r19c-71t9},
  issue     = {24},
  numpages  = {10},
  publisher = {American Physical Society},
  url       = {https://link.aps.org/doi/10.1103/r19c-71t9},
}

@Article{Liu2026,
  author    = {Liu, Luohong and Li, Yuzeng and Wang, Weijia and Zhang, Qicheng and Qiu, Chunyin},
  journal   = {Phys. Rev. Lett.},
  title     = {Observation of Higher-Dimensional Point-Gap Bulk-Boundary Correspondence},
  year      = {2026},
  month     = {Apr},
  pages     = {166602},
  volume    = {136},
  doi       = {10.1103/s4fz-41lg},
  issue     = {16},
  numpages  = {9},
  publisher = {American Physical Society},
  url       = {https://link.aps.org/doi/10.1103/s4fz-41lg},
}

@Article{Mardani2025,
  author    = {Mardani, Yasamin and Pimenta, Rodrigo A. and Sirker, Jesko},
  journal   = {Phys. Rev. B},
  title     = {Exceptional points, bulk-boundary correspondence, and entanglement properties for a dimerized {H}atano-{N}elson model with staggered potentials},
  year      = {2025},
  month     = {Oct},
  pages     = {165133},
  volume    = {112},
  doi       = {10.1103/xfmf-dxdy},
  issue     = {16},
  numpages  = {23},
  publisher = {American Physical Society},
  url       = {https://link.aps.org/doi/10.1103/xfmf-dxdy},
}

@Article{Mong2011,
  author    = {Mong, Roger S. K. and Shivamoggi, Vasudha},
  journal   = {Phys. Rev. B},
  title     = {Edge states and the bulk-boundary correspondence in {D}irac {H}amiltonians},
  year      = {2011},
  month     = {Mar},
  pages     = {125109},
  volume    = {83},
  doi       = {10.1103/PhysRevB.83.125109},
  issue     = {12},
  numpages  = {15},
  publisher = {American Physical Society},
  url       = {https://link.aps.org/doi/10.1103/PhysRevB.83.125109},
}

@Article{Rhim2018,
  author    = {Rhim, Jun-Won and Bardarson, Jens H. and Slager, Robert-Jan},
  journal   = {Phys. Rev. B},
  title     = {Unified bulk-boundary correspondence for band insulators},
  year      = {2018},
  month     = {Mar},
  pages     = {115143},
  volume    = {97},
  doi       = {10.1103/PhysRevB.97.115143},
  issue     = {11},
  numpages  = {20},
  publisher = {American Physical Society},
  url       = {https://link.aps.org/doi/10.1103/PhysRevB.97.115143},
}

@Article{Jin2019,
  author    = {Jin, L. and Song, Z.},
  journal   = {Phys. Rev. B},
  title     = {Bulk-boundary correspondence in a non-{H}ermitian system in one dimension with chiral inversion symmetry},
  year      = {2019},
  month     = {Feb},
  pages     = {081103(R)},
  volume    = {99},
  doi       = {10.1103/PhysRevB.99.081103},
  issue     = {8},
  numpages  = {9},
  publisher = {American Physical Society},
  url       = {https://link.aps.org/doi/10.1103/PhysRevB.99.081103},
}

@Article{Edvardsson2019,
  author    = {Edvardsson, Elisabet and Kunst, Flore K. and Bergholtz, Emil J.},
  journal   = {Phys. Rev. B},
  title     = {Non-{H}ermitian extensions of higher-order topological phases and their biorthogonal bulk-boundary correspondence},
  year      = {2019},
  month     = {Feb},
  pages     = {081302(R)},
  volume    = {99},
  doi       = {10.1103/PhysRevB.99.081302},
  issue     = {8},
  numpages  = {6},
  publisher = {American Physical Society},
  url       = {https://link.aps.org/doi/10.1103/PhysRevB.99.081302},
}

@Article{Hwang2019,
  author    = {Hwang, Yoonseok and Ahn, Junyeong and Yang, Bohm-Jung},
  journal   = {Phys. Rev. B},
  title     = {Fragile topology protected by inversion symmetry: Diagnosis, bulk-boundary correspondence, and Wilson loop},
  year      = {2019},
  month     = {Nov},
  pages     = {205126},
  volume    = {100},
  doi       = {10.1103/PhysRevB.100.205126},
  issue     = {20},
  numpages  = {38},
  publisher = {American Physical Society},
  url       = {https://link.aps.org/doi/10.1103/PhysRevB.100.205126},
}

@Article{Wang2020b,
  author    = {Wang, Xiao-Ran and Guo, Cui-Xian and Kou, Su-Peng},
  journal   = {Phys. Rev. B},
  title     = {Defective edge states and number-anomalous bulk-boundary correspondence in non-{H}ermitian topological systems},
  year      = {2020},
  month     = {Mar},
  pages     = {121116(R)},
  volume    = {101},
  doi       = {10.1103/PhysRevB.101.121116},
  issue     = {12},
  numpages  = {5},
  publisher = {American Physical Society},
  url       = {https://link.aps.org/doi/10.1103/PhysRevB.101.121116},
}

@Article{BayonaPena2025,
  author    = {Bayona-Pena, Pablo and Hanai, Ryo and Mori, Takashi and Hayakawa, Hisao},
  journal   = {Phys. Rev. B},
  title     = {Entanglement spectrum dynamics as a probe for non-{H}ermitian bulk-boundary correspondence in systems with periodic boundaries},
  year      = {2025},
  month     = {Apr},
  pages     = {L140303},
  volume    = {111},
  doi       = {10.1103/PhysRevB.111.L140303},
  issue     = {14},
  numpages  = {6},
  publisher = {American Physical Society},
  url       = {https://link.aps.org/doi/10.1103/PhysRevB.111.L140303},
}

@Article{Trifunovic2019,
  author    = {Trifunovic, Luka and Brouwer, Piet W.},
  journal   = {Phys. Rev. X},
  title     = {Higher-Order Bulk-Boundary Correspondence for Topological Crystalline Phases},
  year      = {2019},
  month     = {Jan},
  pages     = {011012},
  volume    = {9},
  doi       = {10.1103/PhysRevX.9.011012},
  issue     = {1},
  numpages  = {33},
  publisher = {American Physical Society},
  url       = {https://link.aps.org/doi/10.1103/PhysRevX.9.011012},
}

@Article{Yokomizo2019,
  author    = {Yokomizo, Kazuki and Murakami, Shuichi},
  journal   = {Phys. Rev. Lett.},
  title     = {Non-{B}loch Band Theory of Non-{H}ermitian Systems},
  year      = {2019},
  month     = {Aug},
  pages     = {066404},
  volume    = {123},
  doi       = {10.1103/PhysRevLett.123.066404},
  issue     = {6},
  numpages  = {6},
  publisher = {American Physical Society},
  url       = {https://link.aps.org/doi/10.1103/PhysRevLett.123.066404},
}

@Article{Kaneshiro2025,
  author    = {Kaneshiro, Shin and Peters, Robert},
  journal   = {Phys. Rev. B},
  title     = {Symplectic-amoeba formulation of the non-{B}loch band theory for one-dimensional two-band systems},
  year      = {2025},
  month     = {Aug},
  pages     = {075408},
  volume    = {112},
  doi       = {10.1103/5s1z-5t9r},
  issue     = {7},
  numpages  = {10},
  publisher = {American Physical Society},
  url       = {https://link.aps.org/doi/10.1103/5s1z-5t9r},
}

@Article{Kawabata2020,
  author    = {Kawabata, Kohei and Okuma, Nobuyuki and Sato, Masatoshi},
  journal   = {Phys. Rev. B},
  title     = {Non-Bloch band theory of non-{H}ermitian Hamiltonians in the symplectic class},
  year      = {2020},
  month     = {May},
  pages     = {195147},
  volume    = {101},
  doi       = {10.1103/PhysRevB.101.195147},
  issue     = {19},
  numpages  = {12},
  publisher = {American Physical Society},
  url       = {https://link.aps.org/doi/10.1103/PhysRevB.101.195147},
}

@Article{Yokomizo2021a,
  author    = {Yokomizo, Kazuki and Murakami, Shuichi},
  journal   = {Phys. Rev. B},
  title     = {Non-{B}loch band theory in bosonic {B}ogoliubov--de {G}ennes systems},
  year      = {2021},
  month     = {Apr},
  pages     = {165123},
  volume    = {103},
  doi       = {10.1103/PhysRevB.103.165123},
  issue     = {16},
  numpages  = {8},
  publisher = {American Physical Society},
  url       = {https://link.aps.org/doi/10.1103/PhysRevB.103.165123},
}

@Article{Xue2021,
  author    = {Xue, Wen-Tan and Li, Ming-Rui and Hu, Yu-Min and Song, Fei and Wang, Zhong},
  journal   = {Phys. Rev. B},
  title     = {Simple formulas of directional amplification from non-{B}loch band theory},
  year      = {2021},
  month     = {Jun},
  pages     = {L241408},
  volume    = {103},
  doi       = {10.1103/PhysRevB.103.L241408},
  issue     = {24},
  numpages  = {6},
  publisher = {American Physical Society},
  url       = {https://link.aps.org/doi/10.1103/PhysRevB.103.L241408},
}

@Article{Yokomizo2024,
  author    = {Yokomizo, Kazuki and Yoda, Taiki and Ashida, Yuto},
  journal   = {Phys. Rev. B},
  title     = {Non-{B}loch band theory of generalized eigenvalue problems},
  year      = {2024},
  month     = {Mar},
  pages     = {115115},
  volume    = {109},
  doi       = {10.1103/PhysRevB.109.115115},
  issue     = {11},
  numpages  = {9},
  publisher = {American Physical Society},
  url       = {https://link.aps.org/doi/10.1103/PhysRevB.109.115115},
}

@Article{Verma2024,
  author    = {Verma, Sonu and Park, Moon Jip},
  journal   = {Phys. Rev. B},
  title     = {Non-{B}loch band theory of subsymmetry-protected topological phases},
  year      = {2024},
  month     = {Jul},
  pages     = {035424},
  volume    = {110},
  doi       = {10.1103/PhysRevB.110.035424},
  issue     = {3},
  numpages  = {14},
  publisher = {American Physical Society},
  url       = {https://link.aps.org/doi/10.1103/PhysRevB.110.035424},
}

@Article{Hu2024,
  author    = {Hu, Yu-Min and Huang, Yin-Quan and Xue, Wen-Tan and Wang, Zhong},
  journal   = {Phys. Rev. B},
  title     = {Non-{B}loch band theory for non-{H}ermitian continuum systems},
  year      = {2024},
  month     = {Nov},
  pages     = {205429},
  volume    = {110},
  doi       = {10.1103/PhysRevB.110.205429},
  issue     = {20},
  numpages  = {27},
  publisher = {American Physical Society},
  url       = {https://link.aps.org/doi/10.1103/PhysRevB.110.205429},
}

@Article{Wang2024b,
  author    = {Wang, Hong-Yi and Song, Fei and Wang, Zhong},
  journal   = {Phys. Rev. X},
  title     = {Amoeba Formulation of Non-{B}loch Band Theory in Arbitrary Dimensions},
  year      = {2024},
  month     = {Apr},
  pages     = {021011},
  volume    = {14},
  doi       = {10.1103/PhysRevX.14.021011},
  issue     = {2},
  numpages  = {21},
  publisher = {American Physical Society},
  url       = {https://link.aps.org/doi/10.1103/PhysRevX.14.021011},
}

@Article{Song2019a,
  author    = {Song, Fei and Yao, Shunyu and Wang, Zhong},
  journal   = {Phys. Rev. Lett.},
  title     = {Non-{H}ermitian Topological Invariants in Real Space},
  year      = {2019},
  month     = {Dec},
  pages     = {246801},
  volume    = {123},
  doi       = {10.1103/PhysRevLett.123.246801},
  issue     = {24},
  numpages  = {8},
  publisher = {American Physical Society},
  url       = {https://link.aps.org/doi/10.1103/PhysRevLett.123.246801},
}

@Article{Yao2018,
  author    = {Yao, Shunyu and Wang, Zhong},
  journal   = {Phys. Rev. Lett.},
  title     = {Edge States and Topological Invariants of Non-{H}ermitian Systems},
  year      = {2018},
  month     = aug,
  number    = {8},
  pages     = {086803},
  volume    = {121},
  doi       = {10.1103/PhysRevLett.121.086803},
  publisher = {American Physical Society},
  refid     = {10.1103/PhysRevLett.121.086803},
  url       = {https://link.aps.org/doi/10.1103/PhysRevLett.121.086803},
}

@Article{Hodaei2017,
  author  = {Hodaei, Hossein and Hassan, Absar U. and Wittek, Steffen and Garcia-Gracia, Hipolito and El-Ganainy, Ramy and Christodoulides, Demetrios N. and Khajavikhan, Mercedeh},
  journal = {Nature},
  title   = {Enhanced sensitivity at higher-order exceptional points},
  year    = {2017},
  issn    = {1476-4687},
  number  = {7666},
  pages   = {187--191},
  volume  = {548},
  doi     = {10.1038/nature23280},
  refid   = {Hodaei2017},
  url     = {https://doi.org/10.1038/nature23280},
}

@Article{Yuan2023,
  author    = {Yuan, Hao and Zhang, Weixuan and Zhou, Zilong and Wang, Wenlong and Pan, Naiqiao and Feng, Yue and Sun, Houjun and Zhang, Xiangdong},
  journal   = {Adv. Sci.},
  title     = {Non-{H}ermitian Topolectrical Circuit Sensor with High Sensitivity},
  year      = {2023},
  issn      = {2198-3844},
  month     = jul,
  number    = {19},
  pages     = {2301128},
  volume    = {10},
  doi       = {10.1002/advs.202301128},
  publisher = {John Wiley & Sons, Ltd},
  url       = {https://doi.org/10.1002/advs.202301128},
}

@Article{Budich2020,
  author    = {Budich, Jan Carl and Bergholtz, Emil J.},
  journal   = {Phys. Rev. Lett.},
  title     = {Non-{H}ermitian Topological Sensors},
  year      = {2020},
  month     = {Oct},
  pages     = {180403},
  volume    = {125},
  doi       = {10.1103/PhysRevLett.125.180403},
  issue     = {18},
  numpages  = {7},
  publisher = {American Physical Society},
  url       = {https://link.aps.org/doi/10.1103/PhysRevLett.125.180403},
}

@Article{Zhang2019a,
  author    = {Zhang, Mengzhen and Sweeney, William and Hsu, Chia Wei and Yang, Lan and Stone, A. D. and Jiang, Liang},
  journal   = {Phys. Rev. Lett.},
  title     = {Quantum Noise Theory of Exceptional Point Amplifying Sensors},
  year      = {2019},
  month     = {Oct},
  pages     = {180501},
  volume    = {123},
  doi       = {10.1103/PhysRevLett.123.180501},
  issue     = {18},
  numpages  = {6},
  publisher = {American Physical Society},
  url       = {https://link.aps.org/doi/10.1103/PhysRevLett.123.180501},
}

@Article{Wiersig2020,
  author  = {Wiersig, Jan},
  journal = {Nat. Commun.},
  title   = {Prospects and fundamental limits in exceptional point-based sensing},
  year    = {2020},
  issn    = {2041-1723},
  number  = {1},
  pages   = {2454},
  volume  = {11},
  doi     = {10.1038/s41467-020-16373-8},
  refid   = {Wiersig2020},
  url     = {https://doi.org/10.1038/s41467-020-16373-8},
}

@Article{Lau2018,
  author  = {Lau, Hoi-Kwan and Clerk, Aashish A.},
  journal = {Nat. Commun.},
  title   = {Fundamental limits and non-reciprocal approaches in non-{H}ermitian quantum sensing},
  year    = {2018},
  issn    = {2041-1723},
  number  = {1},
  pages   = {4320},
  volume  = {9},
  doi     = {10.1038/s41467-018-06477-7},
  refid   = {Lau2018},
  url     = {https://doi.org/10.1038/s41467-018-06477-7},
}

@Article{Koch2022,
  author    = {Koch, Florian and Budich, Jan Carl},
  journal   = {Phys. Rev. Res.},
  title     = {Quantum non-{H}ermitian topological sensors},
  year      = {2022},
  month     = {Feb},
  pages     = {013113},
  volume    = {4},
  doi       = {10.1103/PhysRevResearch.4.013113},
  issue     = {1},
  numpages  = {8},
  publisher = {American Physical Society},
  url       = {https://link.aps.org/doi/10.1103/PhysRevResearch.4.013113},
}

@Article{Lee2008,
  author    = {Lee, Soo-Young and Ryu, Jung-Wan and Shim, Jeong-Bo and Lee, Sang-Bum and Kim, Sang Wook and An, Kyungwon},
  journal   = {Phys. Rev. A},
  title     = {Divergent {P}etermann factor of interacting resonances in a stadium-shaped microcavity},
  year      = {2008},
  month     = {Jul},
  pages     = {015805},
  volume    = {78},
  doi       = {10.1103/PhysRevA.78.015805},
  issue     = {1},
  numpages  = {4},
  publisher = {American Physical Society},
  url       = {https://link.aps.org/doi/10.1103/PhysRevA.78.015805},
}

@Article{Wang2020c,
  author  = {Wang, Heming and Lai, Yu-Hung and Yuan, Zhiquan and Suh, Myoung-Gyun and Vahala, Kerry},
  journal = {Nat. Commun.},
  title   = {Petermann-factor sensitivity limit near an exceptional point in a {B}rillouin ring laser gyroscope},
  year    = {2020},
  issn    = {2041-1723},
  number  = {1},
  pages   = {1610},
  volume  = {11},
  doi     = {10.1038/s41467-020-15341-6},
  refid   = {Wang2020},
  url     = {https://doi.org/10.1038/s41467-020-15341-6},
}

@Article{Wiersig2020a,
  author    = {Jan Wiersig},
  journal   = {Photon. Res.},
  title     = {Review of exceptional point-based sensors},
  year      = {2020},
  month     = {Sep},
  number    = {9},
  pages     = {1457--1467},
  volume    = {8},
  doi       = {10.1364/PRJ.396115},
  publisher = {Optica Publishing Group},
  url       = {https://opg.optica.org/prj/abstract.cfm?URI=prj-8-9-1457},
}

@Article{Landers2026,
  author  = {Serena Landers and William Tuxbury and Ilya Vitebskiy and Tsampikos Kottos },
  journal = {Sci. Adv.},
  title   = {Noise-resilient exceptional point sensing with immunity to undesired perturbations},
  year    = {2026},
  number  = {9},
  pages   = {eaeb7018},
  volume  = {12},
  doi     = {10.1126/sciadv.aeb7018},
  url     = {https://www.science.org/doi/abs/10.1126/sciadv.aeb7018},
}

@Article{Loughlin2024,
  author    = {Loughlin, Hudson and Sudhir, Vivishek},
  journal   = {Phys. Rev. Lett.},
  title     = {Exceptional-Point Sensors Offer No Fundamental Signal-to-Noise Ratio Enhancement},
  year      = {2024},
  month     = {Jun},
  pages     = {243601},
  volume    = {132},
  doi       = {10.1103/PhysRevLett.132.243601},
  issue     = {24},
  numpages  = {6},
  publisher = {American Physical Society},
  url       = {https://link.aps.org/doi/10.1103/PhysRevLett.132.243601},
}

@Article{Zhong2019,
  author    = {Zhong, Q. and Ren, J. and Khajavikhan, M. and Christodoulides, D. N. and \"Ozdemir, \ifmmode \mbox{\c{S}}\else \c{S}\fi{}. K. and El-Ganainy, R.},
  journal   = {Phys. Rev. Lett.},
  title     = {Sensing with Exceptional Surfaces in Order to Combine Sensitivity with Robustness},
  year      = {2019},
  month     = {Apr},
  pages     = {153902},
  volume    = {122},
  doi       = {10.1103/PhysRevLett.122.153902},
  issue     = {15},
  numpages  = {6},
  publisher = {American Physical Society},
  url       = {https://link.aps.org/doi/10.1103/PhysRevLett.122.153902},
}

@Article{Wang2019a,
  author  = {Wang, Shubo and Hou, Bo and Lu, Weixin and Chen, Yuntian and Zhang, Z. Q. and Chan, C. T.},
  journal = {Nat. Commun.},
  title   = {Arbitrary order exceptional point induced by photonic spin-orbit interaction in coupled resonators},
  year    = {2019},
  issn    = {2041-1723},
  number  = {1},
  pages   = {832},
  volume  = {10},
  doi     = {10.1038/s41467-019-08826-6},
  refid   = {Wang2019},
  url     = {https://doi.org/10.1038/s41467-019-08826-6},
}

@Article{Feng2017,
  author  = {Feng, Liang and El-Ganainy, Ramy and Ge, Li},
  journal = {Nat. Photon.},
  title   = {Non-{H}ermitian photonics based on parity-time symmetry},
  year    = {2017},
  issn    = {1749-4893},
  number  = {12},
  pages   = {752--762},
  volume  = {11},
  doi     = {10.1038/s41566-017-0031-1},
  refid   = {Feng2017},
  url     = {https://doi.org/10.1038/s41566-017-0031-1},
}

@Article{Yuan2023a,
  author   = {Yuan, Hao and Zhang, Weixuan and Zhou, Zilong and Wang, Wenlong and Pan, Naiqiao and Feng, Yue and Sun, Houjun and Zhang, Xiangdong},
  journal  = {Adv. Sci.},
  title    = {Non-{H}ermitian Topolectrical Circuit Sensor with High Sensitivity},
  year     = {2023},
  number   = {19},
  pages    = {2301128},
  volume   = {10},
  doi      = {https://doi.org/10.1002/advs.202301128},
  url      = {https://advanced.onlinelibrary.wiley.com/doi/abs/10.1002/advs.202301128},
}

@Article{Xiao2019,
  author    = {Xiao, Zhicheng and Li, Huanan and Kottos, Tsampikos and Al\`u, Andrea},
  journal   = {Phys. Rev. Lett.},
  title     = {Enhanced Sensing and Nondegraded Thermal Noise Performance Based on $\mathcal{P}\mathcal{T}$-Symmetric Electronic Circuits with a Sixth-Order Exceptional Point},
  year      = {2019},
  month     = {Nov},
  pages     = {213901},
  volume    = {123},
  doi       = {10.1103/PhysRevLett.123.213901},
  issue     = {21},
  numpages  = {6},
  publisher = {American Physical Society},
  url       = {https://link.aps.org/doi/10.1103/PhysRevLett.123.213901},
}

@Article{Zhang2026a,
  author    = {Zhang, Tian and He, Guang-Chen and Chen, Zhao-Xian and Zhang, Xiujuan and Lu, Ming-Hui and Chen, Yan-Feng},
  journal   = {Phys. Rev. B},
  title     = {Noise-robust and highly sensitive sensing empowered by non-{H}ermitian amplification},
  year      = {2026},
  month     = {Mar},
  pages     = {104307},
  volume    = {113},
  doi       = {10.1103/gr8d-b74y},
  issue     = {10},
  numpages  = {10},
  publisher = {American Physical Society},
  url       = {https://link.aps.org/doi/10.1103/gr8d-b74y},
}

@Article{Chen2017a,
  author   = {Chen, Weijian and Kaya {\"O}zdemir, {\c{S}}ahin and Zhao, Guangming and Wiersig, Jan and Yang, Lan},
  journal  = {Nature},
  title    = {Exceptional points enhance sensing in an optical microcavity},
  year     = {2017},
  issn     = {1476-4687},
  number   = {7666},
  pages    = {192--196},
  volume   = {548},
  doi      = {10.1038/nature23281},
  refid    = {Chen2017},
  url      = {https://doi.org/10.1038/nature23281},
}

@Article{Hokmabadi2019,
  author  = {Hokmabadi, Mohammad P. and Schumer, Alexander and Christodoulides, Demetrios N. and Khajavikhan, Mercedeh},
  journal = {Nature},
  title   = {Non-{H}ermitian ring laser gyroscopes with enhanced Sagnac sensitivity},
  year    = {2019},
  issn    = {1476-4687},
  number  = {7785},
  pages   = {70--74},
  volume  = {576},
  doi     = {10.1038/s41586-019-1780-4},
  refid   = {Hokmabadi2019},
  url     = {https://doi.org/10.1038/s41586-019-1780-4},
}

@Article{Mao2023,
  author  = {Xuan Mao and Guo-Qing Qin and Hao Zhang and Bo-Yang Wang and Dan Long and Gui-Qin Li and Gui-Lu Long },
  journal = {Research},
  title   = {Enhanced Sensing Mechanism Based on Shifting an Exceptional Point},
  year    = {2023},
  pages   = {0260},
  volume  = {6},
  doi     = {10.34133/research.0260},
  url     = {https://spj.science.org/doi/abs/10.34133/research.0260},
}

@book{Mehta2004,
  author    = {Mehta, Madan Lal},
  title     = {Random Matrices},
  publisher = {Elsevier},
  year      = {2004},
  edition   = {3},
  isbn      = {9780120884094},
}

@Article{Hatano1996,
  author    = {Hatano, Naomichi and Nelson, David R.},
  journal   = {Phys. Rev. Lett.},
  title     = {Localization Transitions in Non-{H}ermitian Quantum Mechanics},
  year      = {1996},
  month     = {Jul},
  pages     = {570--573},
  volume    = {77},
  doi       = {10.1103/PhysRevLett.77.570},
  issue     = {3},
  numpages  = {0},
  publisher = {American Physical Society},
  url       = {https://link.aps.org/doi/10.1103/PhysRevLett.77.570},
}

@Article{Zhang2020a,
  author    = {Zhang, Kai and Yang, Zhesen and Fang, Chen},
  journal   = {Phys. Rev. Lett.},
  title     = {Correspondence between Winding Numbers and Skin Modes in Non-{H}ermitian Systems},
  year      = {2020},
  month     = {Sep},
  pages     = {126402},
  volume    = {125},
  doi       = {10.1103/PhysRevLett.125.126402},
  issue     = {12},
  numpages  = {6},
  publisher = {American Physical Society},
  url       = {https://link.aps.org/doi/10.1103/PhysRevLett.125.126402},
}

@Article{Song2020,
  author    = {Song, Yiling and Liu, Weiwei and Zheng, Lingzhi and Zhang, Yicong and Wang, Bing and Lu, Peixiang},
  journal   = {Phys. Rev. Appl.},
  title     = {Two-dimensional non-{H}ermitian Skin Effect in a Synthetic Photonic Lattice},
  year      = {2020},
  month     = {Dec},
  pages     = {064076},
  volume    = {14},
  doi       = {10.1103/PhysRevApplied.14.064076},
  issue     = {6},
  numpages  = {9},
  publisher = {American Physical Society},
  url       = {https://link.aps.org/doi/10.1103/PhysRevApplied.14.064076},
}

@Article{Weidemann2020,
  author  = {Sebastian Weidemann and Mark Kremer and Tobias Helbig and Tobias Hofmann and Alexander Stegmaier and Martin Greiter and Ronny Thomale and Alexander Szameit },
  journal = {Science},
  title   = {Topological funneling of light},
  year    = {2020},
  number  = {6488},
  pages   = {311-314},
  volume  = {368},
  doi     = {10.1126/science.aaz8727},
  url     = {https://www.science.org/doi/abs/10.1126/science.aaz8727},
}

@Article{Imhof2018,
  author  = {Imhof, Stefan and Berger, Christian and Bayer, Florian and Brehm, Johannes and Molenkamp, Laurens W. and Kiessling, Tobias and Schindler, Frank and Lee, Ching Hua and Greiter, Martin and Neupert, Titus and Thomale, Ronny},
  journal = {Nat. Phys.},
  title   = {Topolectrical-circuit realization of topological corner modes},
  year    = {2018},
  issn    = {1745-2481},
  number  = {9},
  pages   = {925--929},
  volume  = {14},
  doi     = {10.1038/s41567-018-0246-1},
  refid   = {Imhof2018},
  url     = {https://doi.org/10.1038/s41567-018-0246-1},
}

@book{Kato1966,
  author    = {Kato, Tosio},
  title     = {Perturbation Theory for Linear Operators},
  publisher = {Springer},
  address   = {Berlin},
  year      = {1966}
}

@book{Seyranian2003,
  author    = {Seyranian, Alexander P. and Mailybaev, Alexei A.},
  title     = {Multiparameter Stability Theory with Mechanical Applications},
  publisher = {World Scientific},
  address   = {Singapore},
  year      = {2003}
}

\end{document}